\documentclass{article}

\usepackage{arxiv}

\usepackage[utf8]{inputenc} 
\usepackage[T1]{fontenc}    
\usepackage{hyperref}       
\usepackage{url}            
\usepackage{booktabs}       
\usepackage{amsfonts}       
\usepackage{nicefrac}       
\usepackage{microtype}      
\usepackage{lipsum}

\newcommand{\seams}{AXES}
\usepackage{tikz}
\usepackage{pgfplotstable}
\pgfplotsset{compat=1.7}
\usepgfplotslibrary{groupplots}
\usetikzlibrary{patterns,calc,positioning,fit,external,shapes}
\usepackage{subcaption}
\usepackage{siunitx}

\usepackage[usestackEOL]{stackengine}
\usepackage{multirow}

\usepackage{algorithm} 
\usepackage{algpseudocode} 

\newcommand{\circled}[1]{\raisebox{.5pt}{\textcircled{\raisebox{-.9pt} {#1}}}}

\usepackage{amsmath}

\usepackage{authblk}
\usepackage{fancyhdr}
\usepackage{placeins}

\title{AXES: \underline{A}ppro\underline{x}imation Manager for \underline{E}merging Memory Architecture\underline{s}}

\author[1]{Biswadip~Maity}
\author[1,2]{Bryan~Donyanavard}
\author[2]{Anmol~Surhonne}
\author[1]{Amir~Rahmani}
\author[2]{Andreas~Herkersdorf}
\author[1]{Nikil~Dutt}

\affil[1]{University of California, Irvine, CA \authorcr Email: {\tt \{maityb, bdonyana, a.rahmani, dutt\}@uci.edu}\vspace{1.5ex}}
\affil[2]{Technical University of Munich, Munich, Germany \authorcr Email: {\tt \{anmol.surhonne, herkersdorf\}@tum.de} \vspace{-2ex}} 

\begin{document}
\maketitle

\begin{abstract}
Memory approximation techniques are commonly limited in scope, targeting individual levels of the memory hierarchy.
Existing approximation techniques for a full memory hierarchy determine optimal configurations at design-time provided a goal and application. 
Such policies are rigid: they cannot adapt to unknown workloads and must be redesigned for different memory configurations and technologies. 
We propose \seams{}: the first self-optimizing runtime manager for coordinating configurable approximation knobs across all levels of the memory hierarchy.
\seams{} continuously updates and optimizes its approximation management policy throughout runtime for diverse workloads. 
\seams{} optimizes the approximate memory configuration to minimize power consumption without compromising the quality threshold specified by application developers. 
\seams{} can (1) learn a policy at runtime to manage variable application quality of service (QoS) constraints, (2) automatically optimize for a target metric within those constraints, and (3) coordinate runtime decisions for interdependent knobs and subsystems.
We demonstrate \seams{}' ability to efficiently provide functions 1-3 on a RISC-V Linux platform with approximate memory segments in the on-chip cache and main memory. 
We demonstrate \seams{}' ability to save up to 37\% energy in the memory subsystem without any design-time overhead. We show \seams{}' ability to reduce QoS violations by 75\% with $<5\%$ additional energy.
\end{abstract}


\section{Introduction}
As applications become increasingly resource-intensive, trading off performance and energy in battery-powered systems is crucial. 
Application profiling has revealed memory-accesses as one of the most significant performance, and energy bottlenecks \cite{10.5555/1855840.1855861}. 
Approximate memory is an effective way to alleviate the energy bottleneck in memory for applications that can tolerate output errors caused by inexact memory load/store operations; potentially improving energy consumption, leakage power, latency, or lifetime. 
The inexactness stems from relaxing the need for high-precision storage for some data structures in the application. 

Approximation techniques for different types of memories with configurable degrees of approximation have been previously explored \cite{Shoushtari2015,DBLP:conf/hpca/SmullenMNGS11,Liu2011,10.1145/2540708.2540712,
Qureshi:2009:ELS:1669112.1669117,Cho:2014:ETM:2627369.2627626},
but are typically limited to one or a few levels of the memory hierarchy.
A  holistic solution would need to manage approximation knobs (e.g., $V_{DD}$ for SRAM caches, $t_{REF}$ for DRAM main memory) across the entire memory hierarchy, from cache to main memory, in order to fully exploit memory approximation opportunities.
Furthermore, runtime dynamic reconfiguration of approximation knobs is required to fully leverage the performance/energy tradeoff while honoring application goals, e.g., quality of service (QoS) targets, and system constraints, e.g., minimize energy consumption.

Current approximation techniques using runtime dynamic reconfiguration still depend on the design-time modeling of workloads (application along with input) to determine the optimal knobs of operation for a specific system configuration. 
Such techniques are application-specific and cannot be ported to new systems. 
However, memory technologies are changing very rapidly, and design-time workload profiling dependency introduces significant overhead to utilize approximate memory in emerging memory technologies. 
Furthermore, runtime dynamic reconfiguration across the memory hierarchy requires coordination between multiple knobs (e.g., L1 $V_{DD}$, L2 $V_{DD}$, and DRAM $t_{REF}$).
Coordination is challenging because knobs in one layer can affect other subsystems directly or indirectly (e.g., write errors in L1 affect reads in higher layers).
\begin{figure}[!t]
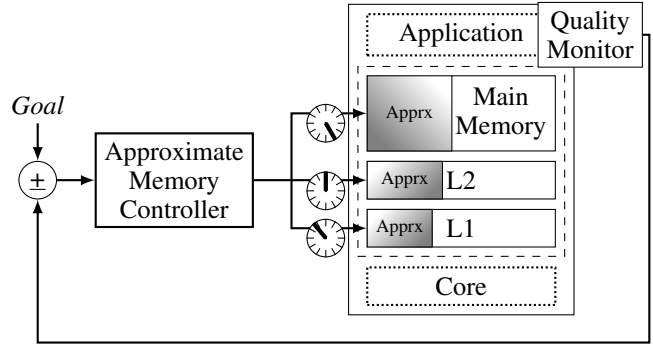

    \centering
    \begin{tikzpicture}[>=latex, every text node part/.style={align=center}]
\def\mwdth{1.25cm}
\def\mhgth{0.5cm}
\def\mdis{0.5cm}

\node [draw, minimum width=\mwdth, minimum height=\mhgth*2, thick] (Controller) at (0,0) {Approximate\\Memory\\Controller} node [below = 0mm of Controller] {};

\node [draw, minimum width=2*\mwdth, minimum height=\mhgth, right = 3*\mdis of Controller] (L2) {L2};
\node [draw, shading = axis, shading angle=225, minimum width=0.8*\mwdth, minimum height=\mhgth, anchor=west] (L2APPRX) at (L2.west) {\scriptsize Apprx};
\node [draw, minimum width=2*\mwdth, minimum height=\mhgth, below = 0.25*\mdis of L2] (L1) {L1};
\node [draw, shading = axis, shading angle=225, minimum width=0.5*\mwdth, minimum height=\mhgth, anchor=west] (L1APPRX) at (L1.west) {\scriptsize Apprx};
\node [draw, thick, densely dotted, minimum width=2*\mwdth, minimum height=\mhgth, below = 0.5*\mdis of L1] (CORE) {Core};
\node [draw, minimum width=2*\mwdth, minimum height=2*\mhgth, above = 0.25*\mdis of L2] (MEM) {} node[anchor=east] (COC) at (MEM.east) {Main\\Memory};
\node [draw, shading = axis, shading angle=135, minimum width=0.9*\mwdth, minimum height=2*\mhgth, anchor=west] (MEMAPPRX) at (MEM.west) {\scriptsize Apprx};
\node[draw, dashed,  fit=(L1) (L2) (MEM)] (MEMSYS) {};
\node [draw, thick, densely dotted, minimum width=2*\mwdth, minimum height=\mhgth, above = 0.5*\mdis of MEM] (APP) {Application};
\node[draw, fit=(MEMSYS) (CORE) (APP)] (SYS) {};
\node [draw, fill=white, minimum width=\mwdth, minimum height=\mhgth, right = -0.5*\mdis of APP] (QM) {Quality\\Monitor};

\node [circle, draw, minimum width=1*\mhgth, minimum height=1*\mhgth, left = \mdis of Controller] (Diff) {};
\node [] () at ($(Diff) - (0,0)$) {+};
\node [] () at ($(Diff) - (0,0.25*\mhgth)$) {$-$};

\draw[->,thick] (Diff) -- (Controller);
\coordinate[right = \mdis of Controller] (ContHelp);
\draw[thick] (Controller) -- (ContHelp) node [midway, below] (Act_Freq) {} ;
\draw[->,thick] (ContHelp) |- (L2) node[near end,below] (L1KNOB) {}; \node [] () at (L1KNOB) {\input{figs/knob1}} ;
\draw[->,thick] (ContHelp) |- (L1) node[near end,below] (L2KNOB) {}; \node [] () at (L2KNOB) {\input{figs/knob2}} ;
\draw[->,thick] (ContHelp) |- (MEM) node[near end,below] (MEMKNOB) {}; \node [] () at (MEMKNOB) {\input{figs/knob3}} ;

\node [above= \mdis of Diff, opacity=0.0] (tmp) {Goal};
\node (Ref) at (tmp) {\textit{Goal}};
\draw[->,thick] (Ref) -- (Diff);

\coordinate[] (FBhelp) at ($(CORE) + (5*\mdis, -1.5*\mdis)$);
\draw[->,thick] (QM) -| node [near start, above] (Sen_Heart) {} node [near start, below] (Sen_IPS) {} (FBhelp) -| (Diff);

\end{tikzpicture}
    \caption{Runtime management of approximation knobs using output quality monitoring.}
    \label{fig:single-knob}
\end{figure}


Consider the system shown in Figure \ref{fig:single-knob}. 
The memory hierarchy (in this case: L1 cache, L2 cache, and main memory) exposes tunable knobs (e.g., operating voltage for L1 and L2, and data refresh period for main memory) that control the degree of approximation. 
Each knob introduces a new degree of freedom and increases the configuration space exponentially. 
Furthermore, satisfying even a single-objective function poses non-trivial optimization challenges, with an additional level of complexity arising from the optimization of multiple objective functions to determine the optimal system configuration.
Researchers have proposed frameworks for exploring the configuration space at design-time and determining static optimal knob settings for an approximate memory hierarchy before deployment \cite{Yarmand2020, Jiao2020}. 
More flexible solutions have been proposed to provide dynamic configuration of knobs at runtime but require identifying workload-specific system dynamics at design-time \cite{Teimoori2018, Maity2019, cecsEsl}.
Apriori knowledge limits the ability to adapt to changing workloads, and further assumes that the system and workload are observable ahead of deployment.
On the other hand, determining the optimal knob configuration for unknown applications and new inputs at runtime is an extremely challenging decision process. 


To address these challenges, we develop \seams{}, a model-free method to tune memory knobs without any previous knowledge about the system (memory-hierarchy as well as the workload). 
\seams{} eases the design of systems with approximate memory by enabling deployment without going through design time exploration of configuration knobs. 
\seams{}' methodology is independent of the underlying memory technologies and works regardless of the nature of knobs.
Once deployed, \seams{} can learn the optimal knob configuration for unknown applications, resulting in self-optimizing systems.



\usetikzlibrary{backgrounds}

\newcommand{\motivationpicA}[3]{
\begin{tikzpicture}
    \begin{axis}[
        point meta min=0,
        point meta max=#2,
        xlabel={L1 DCache Vdd level},
        ylabel={L2 DCache Vdd level},
        xtick={0.3,0.35,0.4,0.45,0.5,0.55},
        ytick={0.3,0.35,0.4,0.45,0.5,0.55},
        width=5cm,
        height=5cm
    ]
    \begin{scope}[on background layer]
    \addplot[
        scatter,
        only marks,
        scatter src=explicit,
        mark=*,
        scatter/use mapped color={
            draw=mapped color,
            fill=mapped color,
        },
        visualization depends on={\thisrow{C} \as \perpointmarksize},
        scatter/@pre marker code/.append style={
            /tikz/mark size={1pt+abs(\perpointmarksize/#3)}
        },
    ] table [meta=C, col sep=comma] {#1};
    \end{scope}
    \node[mark size=7pt,color=black, fill opacity=0.04] at (100,0) {\pgfuseplotmark{*}};
    \node[mark size=7pt,color=black, fill opacity=0.04] at (50,50) {\pgfuseplotmark{*}};
    \end{axis}
\end{tikzpicture}
}

\newcommand{\motivationpicB}[3]{
\begin{tikzpicture}
    \begin{axis}[
        point meta min=0,
        point meta max=#2,
        xlabel={L1 DCache Vdd level},
        ylabel={L2 DCache Vdd level},
        xtick={0.3,0.35,0.4,0.45,0.5,0.55},
        ytick={0.3,0.35,0.4,0.45,0.5,0.55},
        width=5cm,
        height=5cm
    ]
    \begin{scope}[on background layer]
    \addplot[
        scatter,
        only marks,
        scatter src=explicit,
        mark=*,
        scatter/use mapped color={
            draw=mapped color,
            fill=mapped color,
        },
        visualization depends on={\thisrow{C} \as \perpointmarksize},
        scatter/@pre marker code/.append style={
            /tikz/mark size={1pt+abs(\perpointmarksize/#3)}
        },
    ] table [meta=C, col sep=comma] {#1};
    \end{scope}
    \node[mark size=7pt,color=black, fill opacity=0.04] at (150,0) {\pgfuseplotmark{*}};
    \node[mark size=7pt,color=black, fill opacity=0.04] at (50,50) {\pgfuseplotmark{*}};
    \node[mark size=7pt,color=black, fill opacity=0.04] at (50,100) {\pgfuseplotmark{*}};
    \end{axis}
\end{tikzpicture}
}

\newcommand{\motivationpicwithbar}[3]{
\begin{tikzpicture}
    \begin{axis}[
        colorbar,
        colorbar style={
        ytick={0,4,...,24},
        },
        point meta min=0,
        point meta max=#2,
        xlabel={L1 DCache Vdd level},
        ylabel={L2 DCache Vdd level},
        xtick={0.3,0.35,0.4,0.45,0.5,0.55},
        ytick={0.3,0.35,0.4,0.45,0.5,0.55},
        width=5cm,
        height=5cm
    ]
    \begin{scope}[on background layer]
        \addplot[
            scatter,
            only marks,
            scatter src=explicit,
            mark=*,
            scatter/use mapped color={
                draw=mapped color,
                fill=mapped color,
            },
            visualization depends on={\thisrow{C} \as \perpointmarksize},
            scatter/@pre marker code/.append style={
                /tikz/mark size={1pt+abs(\perpointmarksize/#3)}
            },
        ] table [meta=C, col sep=comma] {#1};
    \end{scope}
    \node[mark size=7pt,color=black, fill opacity=0.04] at (100,0) {\pgfuseplotmark{*}};
    \node[mark size=7pt,color=black, fill opacity=0.04] at (150,0) {\pgfuseplotmark{*}};
    \node[mark size=7pt,color=black, fill opacity=0.04] at (200,0) {\pgfuseplotmark{*}};
    \node[mark size=7pt,color=black, fill opacity=0.04] at (0,100) {\pgfuseplotmark{*}};
    \node[mark size=7pt,color=black, fill opacity=0.04] at (0,150) {\pgfuseplotmark{*}};
    \node[mark size=7pt,color=black, fill opacity=0.04] at (0,200) {\pgfuseplotmark{*}};
    \end{axis}
\end{tikzpicture}
}


\textbf{The main contributions of this paper are as follows:}
\begin{enumerate}
    \item Enable \textbf{self-optimization} of multi-level memory approximation knobs through \seams{}, a runtime resource manager using reinforcement learning. Self-optimization is demonstrated by finding the system configuration for unknown workloads at runtime, as well as the dynamic management of quality of service (QoS).
    \item Enable \textbf{coordination} between multiple memory system knobs without explicit communication. Coordination is demonstrated through dynamic runtime reconfiguration of multiple knobs by continuously evaluating different subsystem configurations (e.g., $\uparrow$ L1 knob, $\downarrow$ L2 knob, vs $\downarrow$ L1 knob, $\uparrow$ L2 knob).
    \item An approximate memory management approach that is (a) technology agnostic, (b) application-independent, and (c) can easily be applied to any multi-level memory hierarchy. 
    \item \textbf{Experimental case study}: A software implementation of \seams{} is evaluated using an FPGA board with a modified RISC-V processing core to validate the approach.
\end{enumerate}
We believe \seams{} will enable quick adoption of approximation using a variety of memory nodes.
\section{Motivation}
The QoS delivered by a given configuration of approximation knobs varies widely based on the application and current input. 
Even for a fixed workload (application and input), the configuration space grows exponentially with each additional knob (e.g., one knob = 4 states, two knobs = 16 states, three knobs = 64 states). 
More examples of approximation knobs for different memory technologies are presented in Table \ref{tab:tech}. In a memory hierarchy, knobs are at least partially interdependent: changing one knob affects multiple subsystems in ways that are complex to predict. (e.g., changing L1 $V_{DD}$ introduces errors in L1, which subsequently propagate to L2.)
This makes the configuration problem extremely challenging. 

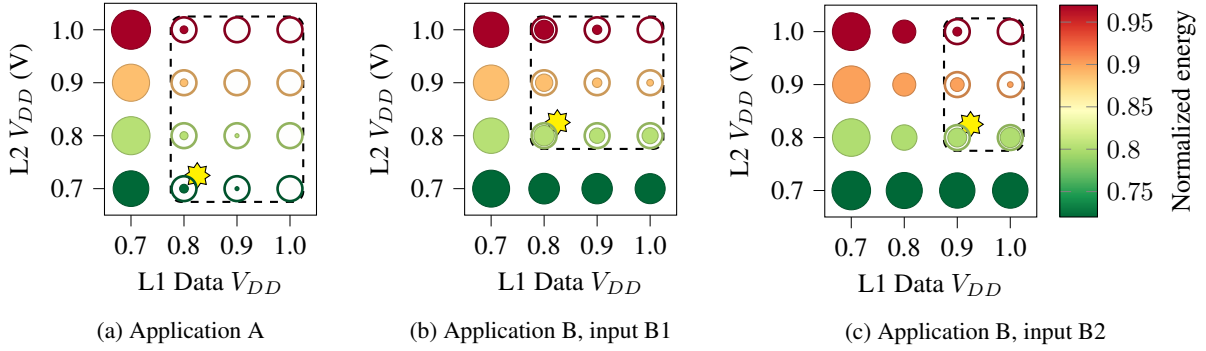
\begin{figure*}
    \begin{subfigure}[b!]{0.29\linewidth}
\begin{tikzpicture}

\begin{axis}[
height=4.4cm,
minor xtick={},
minor ytick={},
tick align=outside,
tick pos=left,
width=4.4cm,
x grid style={white!69.0196078431373!black},
xlabel={L1 Data \(\displaystyle V_{DD}\)},
xmin=0.65, xmax=1.05,
xtick style={color=black},
xtick={0.7,0.8,0.9,1},
xticklabels={0.7,0.8,0.9,1.0},
y grid style={white!69.0196078431373!black},
ylabel={L2 \(\displaystyle V_{DD}\) (V)},
ymin=0.65, ymax=1.05,
ytick style={color=black},
ytick={0.7,0.8,0.9,1},
yticklabels={0.7,0.8,0.9,1.0}
]
\addplot [only marks, scatter,colormap={mymap}{[1pt]
  rgb(0pt)=(0,0.407843137254902,0.215686274509804);
  rgb(1pt)=(0.101960784313725,0.596078431372549,0.313725490196078);
  rgb(2pt)=(0.4,0.741176470588235,0.388235294117647);
  rgb(3pt)=(0.650980392156863,0.850980392156863,0.415686274509804);
  rgb(4pt)=(0.850980392156863,0.937254901960784,0.545098039215686);
  rgb(5pt)=(1,1,0.749019607843137);
  rgb(6pt)=(0.996078431372549,0.87843137254902,0.545098039215686);
  rgb(7pt)=(0.992156862745098,0.682352941176471,0.380392156862745);
  rgb(8pt)=(0.956862745098039,0.427450980392157,0.262745098039216);
  rgb(9pt)=(0.843137254901961,0.188235294117647,0.152941176470588);
  rgb(10pt)=(0.647058823529412,0,0.149019607843137)
}, line width=1pt, fill opacity=0.01, visualization depends on={\thisrow{sizedata} \as\perpointmarksize}, scatter/@pre marker code/.append style={/tikz/mark size=\perpointmarksize}]
table{%
x                      y                      colordata              sizedata
0.8 0.7 0.85162 4.7
0.8 0.8 0.88012 4.7
0.8 0.9 0.91242 4.7
0.8 1 0.94852 4.7
0.9 0.7 0.87593 4.7
0.9 0.8 0.90443 4.7
0.9 0.9 0.93673 4.7
0.9 1 0.97283 4.7
1 0.7 0.9031 4.7
1 0.8 0.9316 4.7
1 0.9 0.9639 4.7
1 1 1 4.7
};
\addplot [only marks, scatter, colormap={mymap}{[1pt]
  rgb(0pt)=(0,0.407843137254902,0.215686274509804);
  rgb(1pt)=(0.101960784313725,0.596078431372549,0.313725490196078);
  rgb(2pt)=(0.4,0.741176470588235,0.388235294117647);
  rgb(3pt)=(0.650980392156863,0.850980392156863,0.415686274509804);
  rgb(4pt)=(0.850980392156863,0.937254901960784,0.545098039215686);
  rgb(5pt)=(1,1,0.749019607843137);
  rgb(6pt)=(0.996078431372549,0.87843137254902,0.545098039215686);
  rgb(7pt)=(0.992156862745098,0.682352941176471,0.380392156862745);
  rgb(8pt)=(0.956862745098039,0.427450980392157,0.262745098039216);
  rgb(9pt)=(0.843137254901961,0.188235294117647,0.152941176470588);
  rgb(10pt)=(0.647058823529412,0,0.149019607843137)
}, visualization depends on={\thisrow{sizedata} \as\perpointmarksize}, scatter/@pre marker code/.append style={/tikz/mark size=\perpointmarksize}]
table [x=x, y=y, meta=colordata]{%
x                      y                      colordata              sizedata
0.7 0.7 0.83017 6.74252582972741
0.7 0.8 0.85867 7.20229493231534
0.7 0.9 0.89097 7.02478338848061
0.7 1 0.92707 7.38868675473589
0.8 0.7 0.85162 1.6491330048102
0.8 0.8 0.88012 1.52686004807582
0.8 0.9 0.91242 1.46189348990888
0.8 1 0.94852 1.46189348990888
0.9 0.7 0.87593 0.76363846909955
0.9 0.8 0.90443 0.76363846909955
0.9 0.9 0.93673 0
0.9 1 0.97283 0
1 0.7 0.9031 0
1 0.8 0.9316 0
1 0.9 0.9639 0
1 1 1 0
};
    \draw[dashed,thick, rounded corners] (axis cs:0.775,1.025) rectangle (axis cs:1.025,0.675);
    \node[draw,star,fill=yellow,star points=8, inner sep=0.8mm] (OPT) at (axis cs:0.825,0.725) {};
\end{axis}
\end{tikzpicture}
        \caption{Application A}
        \label{subfig:CombinationLeft}
    \end{subfigure}%
    \begin{subfigure}[b!]{.29\linewidth}
\begin{tikzpicture}

\begin{axis}[
height=4.4cm,
minor xtick={},
minor ytick={},
tick align=outside,
tick pos=left,
width=4.4cm,
x grid style={white!69.0196078431373!black},
xlabel={L1 Data \(\displaystyle V_{DD}\)},
xmin=0.65, xmax=1.05,
xtick style={color=black},
xtick={0.7,0.8,0.9,1},
xticklabels={0.7,0.8,0.9,1.0},
y grid style={white!69.0196078431373!black},
ylabel={L2 \(\displaystyle V_{DD}\) (V)},
ymin=0.65, ymax=1.05,
ytick style={color=black},
ytick={0.7,0.8,0.9,1},
yticklabels={0.7,0.8,0.9,1.0}
]
\addplot [only marks, scatter,colormap={mymap}{[1pt]
  rgb(0pt)=(0,0.407843137254902,0.215686274509804);
  rgb(1pt)=(0.101960784313725,0.596078431372549,0.313725490196078);
  rgb(2pt)=(0.4,0.741176470588235,0.388235294117647);
  rgb(3pt)=(0.650980392156863,0.850980392156863,0.415686274509804);
  rgb(4pt)=(0.850980392156863,0.937254901960784,0.545098039215686);
  rgb(5pt)=(1,1,0.749019607843137);
  rgb(6pt)=(0.996078431372549,0.87843137254902,0.545098039215686);
  rgb(7pt)=(0.992156862745098,0.682352941176471,0.380392156862745);
  rgb(8pt)=(0.956862745098039,0.427450980392157,0.262745098039216);
  rgb(9pt)=(0.843137254901961,0.188235294117647,0.152941176470588);
  rgb(10pt)=(0.647058823529412,0,0.149019607843137)
}, line width=1pt, fill opacity=0.01, visualization depends on={\thisrow{sizedata} \as\perpointmarksize}, scatter/@pre marker code/.append style={/tikz/mark size=\perpointmarksize}]
table[x=x, y=y, meta=colordata]{%
x                      y                      colordata              sizedata
0.8 0.8 0.777115843054648 4.7
0.8 0.9 0.837533848764923 4.7
0.8 1 0.905059855146995 4.7
0.9 0.8 0.802142640355471 4.7
0.9 0.9 0.862560646065747 4.7
0.9 1 0.930086652447819 4.7
1 0.8 0.830113766750509 4.7
1 0.9 0.890531772460785 4.7
1 1 0.958057778842857 4.7
};
\addplot [only marks, scatter,colormap={mymap}{[1pt]
  rgb(0pt)=(0,0.407843137254902,0.215686274509804);
  rgb(1pt)=(0.101960784313725,0.596078431372549,0.313725490196078);
  rgb(2pt)=(0.4,0.741176470588235,0.388235294117647);
  rgb(3pt)=(0.650980392156863,0.850980392156863,0.415686274509804);
  rgb(4pt)=(0.850980392156863,0.937254901960784,0.545098039215686);
  rgb(5pt)=(1,1,0.749019607843137);
  rgb(6pt)=(0.996078431372549,0.87843137254902,0.545098039215686);
  rgb(7pt)=(0.992156862745098,0.682352941176471,0.380392156862745);
  rgb(8pt)=(0.956862745098039,0.427450980392157,0.262745098039216);
  rgb(9pt)=(0.843137254901961,0.188235294117647,0.152941176470588);
  rgb(10pt)=(0.647058823529412,0,0.149019607843137)
}, visualization depends on={\thisrow{sizedata} \as\perpointmarksize}, scatter/@pre marker code/.append style={/tikz/mark size=\perpointmarksize}]
table [x=x, y=y, meta=colordata]{%
x                      y                      colordata              sizedata
0.7 0.7 0.701723369809561 6.98535333568136
0.7 0.8 0.755033374848039 6.7048257870554
0.7 0.9 0.815451380558314 6.71042481138034
0.7 1 0.882977386940387 6.66303828863736
0.8 0.7 0.72380583801617 5.78657023034887
0.8 0.8 0.777115843054648 3.86566891335638
0.8 0.9 0.837533848764923 3.27956593432449
0.8 1 0.905059855146995 3.58442745552028
0.9 0.7 0.748832635316993 5.74931642280829
0.9 0.8 0.802142640355471 2.90511952082308
0.9 0.9 0.862560646065747 1.7844012370561
0.9 1 0.930086652447819 1.76107071375479
1 0.7 0.776803761712031 5.68108385266184
1 0.8 0.830113766750509 3.03643019886317
1 0.9 0.890531772460785 1.26977380619487
1 1 0.958057778842857 0
};
    \draw[dashed,thick, rounded corners] (axis cs:0.775,1.025) rectangle (axis cs:1.025,0.775);
    \node[draw,star,fill=yellow,star points=8, inner sep=0.8mm] (OPT) at (axis cs:0.825,0.825) {};
\end{axis}
\end{tikzpicture}
        \caption{Application B, input B1}
        \label{subfig:CombinationRight}
    \end{subfigure}%
    \begin{subfigure}[b!]{.41\linewidth}
\begin{tikzpicture}

\begin{axis}[
colorbar,
colorbar style={ytick={0.75,0.8,0.85,0.9,0.95},minor ytick={},ylabel near ticks,ylabel={Normalized energy},
},
colormap={mymap}{[1pt]
 rgb(0pt)=(0,0.407843137254902,0.215686274509804);
 rgb(1pt)=(0.101960784313725,0.596078431372549,0.313725490196078);
 rgb(2pt)=(0.4,0.741176470588235,0.388235294117647);
 rgb(3pt)=(0.650980392156863,0.850980392156863,0.415686274509804);
 rgb(4pt)=(0.850980392156863,0.937254901960784,0.545098039215686);
 rgb(5pt)=(1,1,0.749019607843137);
 rgb(6pt)=(0.996078431372549,0.87843137254902,0.545098039215686);
 rgb(7pt)=(0.992156862745098,0.682352941176471,0.380392156862745);
 rgb(8pt)=(0.956862745098039,0.427450980392157,0.262745098039216);
rgb(9pt)=(0.843137254901961,0.188235294117647,0.152941176470588);
 rgb(10pt)=(0.647058823529412,0,0.149019607843137)
},
height=4.4cm,
minor xtick={},
minor ytick={},
point meta max=0.97,
point meta min=0.72,
tick align=outside,
tick pos=left,
width=4.4cm,
x grid style={white!69.0196078431373!black},
xlabel={L1 Data \(\displaystyle V_{DD}\)},
xmin=0.65, xmax=1.05,
xtick style={color=black},
xtick={0.7,0.8,0.9,1},
xticklabels={0.7,0.8,0.9,1.0},
y grid style={white!69.0196078431373!black},
ylabel={L2 \(\displaystyle V_{DD}\) (V)},
ymin=0.65, ymax=1.05,
ytick style={color=black},
ytick={0.7,0.8,0.9,1},
yticklabels={0.7,0.8,0.9,1.0}
]
\addplot [only marks, scatter,colormap={mymap}{[1pt]
  rgb(0pt)=(0,0.407843137254902,0.215686274509804);
  rgb(1pt)=(0.101960784313725,0.596078431372549,0.313725490196078);
  rgb(2pt)=(0.4,0.741176470588235,0.388235294117647);
  rgb(3pt)=(0.650980392156863,0.850980392156863,0.415686274509804);
  rgb(4pt)=(0.850980392156863,0.937254901960784,0.545098039215686);
  rgb(5pt)=(1,1,0.749019607843137);
  rgb(6pt)=(0.996078431372549,0.87843137254902,0.545098039215686);
  rgb(7pt)=(0.992156862745098,0.682352941176471,0.380392156862745);
  rgb(8pt)=(0.956862745098039,0.427450980392157,0.262745098039216);
  rgb(9pt)=(0.843137254901961,0.188235294117647,0.152941176470588);
  rgb(10pt)=(0.647058823529412,0,0.149019607843137)
}, line width=1pt, fill opacity=0.01, visualization depends on={\thisrow{sizedata} \as\perpointmarksize}, scatter/@pre marker code/.append style={/tikz/mark size=\perpointmarksize}]
table [x=x, y=y, meta=colordata]{%
x                      y                      colordata              sizedata
0.9 0.8 0.802123469763578 4.7
0.9 0.9 0.862400107779215 4.7
0.9 1 0.929768114973162 4.7
1 0.8 0.830214938217802 4.7
1 0.9 0.890491576233439 4.7
1 1 0.957859583427386 4.7
};
\addplot [only marks, scatter, colormap={mymap}{[1pt]
  rgb(0pt)=(0,0.407843137254902,0.215686274509804);
  rgb(1pt)=(0.101960784313725,0.596078431372549,0.313725490196078);
  rgb(2pt)=(0.4,0.741176470588235,0.388235294117647);
  rgb(3pt)=(0.650980392156863,0.850980392156863,0.415686274509804);
  rgb(4pt)=(0.850980392156863,0.937254901960784,0.545098039215686);
  rgb(5pt)=(1,1,0.749019607843137);
  rgb(6pt)=(0.996078431372549,0.87843137254902,0.545098039215686);
  rgb(7pt)=(0.992156862745098,0.682352941176471,0.380392156862745);
  rgb(8pt)=(0.956862745098039,0.427450980392157,0.262745098039216);
  rgb(9pt)=(0.843137254901961,0.188235294117647,0.152941176470588);
  rgb(10pt)=(0.647058823529412,0,0.149019607843137)
}, visualization depends on={\thisrow{sizedata} \as\perpointmarksize}, scatter/@pre marker code/.append style={/tikz/mark size=\perpointmarksize}]
table [x=x, y=y, meta=colordata]{%
x                      y                      colordata              sizedata
0.7 0.7 0.70162625405598 7.29864508737139
0.7 0.8 0.754811522893307 7.1153389407624
0.7 0.9 0.815088160908943 7.07108842027531
0.7 1 0.88245616810289 7.13438570139423
0.8 0.7 0.72380372915142 6.81506059295014
0.8 0.8 0.776988997988747 4.80585543894069
0.8 0.9 0.837265636004383 4.3512840064822
0.8 1 0.90463364319833 4.36931511644595
0.9 0.7 0.748938200926252 6.68207042202004
0.9 0.8 0.802123469763578 3.5730841884392
0.9 0.9 0.862400107779215 2.60447192609298
0.9 1 0.929768114973162 1.67111630866084
1 0.7 0.777029669380476 6.70439451147429
1 0.8 0.830214938217802 3.80578935969337
1 0.9 0.890491576233439 1.06774824181233
1 1 0.957859583427386 0
};
    \draw[dashed,thick, rounded corners] (axis cs:0.875,1.025) rectangle (axis cs:1.025,0.775);
    \node[draw,star,fill=yellow,star points=8, inner sep=0.8mm] (OPT) at (axis cs:0.925,0.825) {};
    
\end{axis}
\end{tikzpicture}
        \caption{Application B, input B2}
        \label{subfig:CombinationRightRight}
    \end{subfigure}
    \caption{Effect of configuration knobs on cache layers (L1 data cache and L2 shared cache) for two different applications (A and B), and different inputs within an application (B1 and B2 within B). The dot diameter indicates the number of errors (smaller is better: no dot means no errors), and the color indicates normalized total energy usage (normalized to 1V:1V case). The outer circle represents the quality-constraint which the system must meet. For knob configurations where there is no outer circle, the system fails to meet the quality constraint. Feasible operating regions that can achieve the target QoS are outlined in dashed rectangles, and the optimal setting is indicated by a star.}
    \label{fig:Combination}
\end{figure*}

\subsection{Case Study}
To illustrate this challenge, consider a system equipped with an approximate memory subsystem, as shown in Figure \ref{fig:single-knob}. 
The application's source code is annotated with a quality monitor and is running on a system that supports approximate memory. 
The approximate memory subsystem consists of three layers of hierarchy, including an SRAM L1 cache memory, an SRAM L2 cache memory, and a DRAM main memory. 
These memories have an `exact' and `approximate' region in which application data can be mapped.  
The degree of approximation varies based on the memory technology: in this work, the voltage level for SRAM cache and refresh rate for DRAM main memory. 
Approximation can be controlled at each layer of the memory hierarchy, and the knob setting impacts the application QoS measured by the quality monitor.
\texttt{malloc} calls from the application to the Linux kernel are modified by the developer to indicate which data can be mapped to approximate regions. 
The complete experimental setup is described in Section \ref{experimentalSetup}.

Figure \ref{fig:Combination} is an illustrative example showing variations in QoS observed across different configurations of L1 and L2 approximation knobs, as well as across different applications. 
The DRAM knob is fixed for the sake of simplicity. 
The dots' size represents the QoS (i.e., number of errors, smaller is better). 
We observe the effect of configuration knobs on two applications:
\begin{itemize}
    \item \textbf{Application A}: A memory write-read  kernel that writes 512 64-bit numbers in the main memory and then reads the numbers from main memory. The QoS metric for this kernel is defined as the total number of bit flips that occur during the write-read cycle. The QoS and average energy for each knob configuration is shown in Figure \ref{subfig:CombinationLeft}. 
    \item \textbf{Application B}: The Canny-edge detection application as described in Section \ref{experimentalSetup_app}. The QoS metric for this application is the \texttt{rmse} (Root Mean Square Error) between the pixels of the approximate runs and the exact runs of the application. The QoS and average energy for knob configurations corresponding to two different inputs (i.e., scenes), B1 and B2, is shown in Figures \ref{subfig:CombinationRight} and \ref{subfig:CombinationRightRight} respectively.
\end{itemize}
\begin{scriptsize}
    \begin{table*}[tbh]
        \caption{Examples of approximate memory technology knobs.}
        \label{tab:tech}
        \centering
        \resizebox{\linewidth}{!}{
            \begin{tabular}{|l|c|c|c|c|c|}
                \hline
                                       & \textbf{Technology}       & \textbf{Memory Type}         & \textbf{Technology Knobs}    & \textbf{Knob objective} & \textbf{Reference}                                                   \\
                \hline
                \hline
                \multirow{2}{*}{Cache} & SRAM                      & Volatile                     & Operating Voltage ($V_{DD}$) & Energy savings          & ASPLOS'12\cite{10.1145/2150976.2151008}, ESL'15\cite{Shoushtari2015} \\ \cline{2-6}
                                       & STT-RAM                   & Non-Volatile                 &
                \begin{tabular}[c]{@{}c@{}}
                    Read Voltage ($V_{read}$) \\
                    Write Pulse Duration ($t_{write}$)
                \end{tabular}
                                       & Energy savings            &
                \begin{tabular}[c]{@{}c@{}}
                    HPCA'11\cite{DBLP:conf/hpca/SmullenMNGS11}, CASES'15\cite{7324548} \\ ISLPED'17\cite{8009198}\end{tabular}
                \\
                \hline
                \multirow{2}{*}{
                    \begin{tabular}[c]{@{}c@{}}
                        Main \\
                        Memory
                    \end{tabular}}
                                       & DRAM                      & Volatile                     & \begin{tabular}[c]{@{}c@{}}Data Refresh Period ($t_{REF}$)\\ Operating Voltage ($V_{DD}$)\\ Row Activation Delay ($t_{RCD}$)\end{tabular}
                                       & \begin{tabular}[c]{@{}c@{}}
                    Energy savings \\
                    Reduce Latency
                \end{tabular}
                                       & \begin{tabular}[c]{@{}c@{}}
                    ASPLOS'11\cite{Liu2011},                     \\
                    ISLPED'14\cite{Cho:2014:ETM:2627369.2627626} \\
                    MICRO'19\cite{10.1145/3352460.3358280}
                \end{tabular}
                \\ \cline{2-6}
                                       & PCM
                                       & Non-Volatile              & Data Comparison Write ($Th$)
                                       & \begin{tabular}[c]{@{}c@{}}
                    Energy savings \\
                    Increase lifetime
                \end{tabular}
                                       &
                MICRO'09\cite{Qureshi:2009:ELS:1669112.1669117}, MICRO'13\cite{10.1145/2540708.2540712}
                \\ \hline
            \end{tabular}
        }
    \end{table*}
\end{scriptsize}
We make the following key observations: 
First, we observe that  configurations achieving a target QoS vary both \emph{within and between} applications.
In Figure~\ref{fig:Combination}, we define a feasible region (dashed rectangle) by identifying the set of configurations that achieve acceptable QoS.
Depending on the workload, the feasible regions of operation are different.
The difference is seen in the varying bounding boxes of Figure \ref{subfig:CombinationLeft} (Application A), and Figure \ref{subfig:CombinationRight} (Application B). 
Even within the same application, the acceptable regions of operation vary based on the dynamic inputs to the application at runtime as seen in Figure \ref{subfig:CombinationRight} (input B1) and Figure \ref{subfig:CombinationRightRight} (input B2).

Second, we observe that even within the feasible regions, the achieved QoS varies across applications and inputs.
In some cases, the outer circle and inner circle are well separated, implying that there is still room for approximation.
However, in some cases, the inner circle is very close to the outer circle, implying that the QoS is reaching its threshold.

Third, we observe that even the same configuration of knobs (e.g., L1:\SI{0.7}{\volt}, L2:\SI{0.9}{\volt}) have different power characteristics with respect to different applications and different inputs within an application. 
This results in varying optimal design points (marked with a star).

This simple example demonstrates that even for the same memory technology, it is hard to predict the resulting QoS and energy when  knobs are changed in only two layers of the memory hierarchy;
i.e., the dynamics between the system and application vary both within and between applications.
We expect that finding the optimal configuration for additional layers of a memory hierarchy or new memory technologies will only exacerbate these challenges, with current state-of-the-art techniques (summarized next) insufficient for determining the complex interactions of knob configurations 
for multi-level approximate memories.

\subsection{
Approximation: State-of-the-art}
\subsubsection{Open-loop Control}
A popular approximation strategy is to use design-time techniques to find optimal knob configurations
\cite{10.1109/MICRO.2014.22, 10.1145/3352460.3358280, 10.5555/2665671.2665746, DBLP:conf/hpca/MoreauWNSECO15}.
Based on the application profile, approximation knobs are determined before deployment and are expected to meet the QoS requirements throughout the application's lifetime. 
For an open-loop system 
designers must design the system with the worst-case scenario in mind and are unable to exploit the full potential of the approximation knobs at runtime. 
Additionally, application programmers are burdened with the task of setting memory approximation knobs through intensive profiling of the target workloads at design time\cite{10.1145/3352460.3358280, Yarmand2020, Teimoori2018}.
Thus, open-loop control techniques are application-specific and not portable to new systems.

\subsubsection{Model-based Closed-loop Control}
To address the lack of reconfiguration in open-loop systems, state-of-the-art alternatives reconfigure approximation knobs using closed-loop controllers.
\cite{10.1145/3352460.3358298,
10.1145/2694344.2694365,
10.1145/3358203,
conf/hpca/GrigorianFR15}
The controllers are generated based on a system model identified at design-time.
Closed-loop control aims to alleviate the programmer's burden at design-time by using feedback at runtime. 
Design-time models consider the difficulty of specifying an under-designed memory's parameters by measuring the output accuracy in different settings. 
However, with the number of system parameters on the rise, system identification is becoming impractical for capturing the effects of one knob on another. 
Coordination in control theory requires a formal Multiple-input-multiple-output (MIMO) method, but designing a MIMO controller requires nontrivial design-time effort. 
Additionally, such models are rigid: models must be generated for each memory technology, with an underlying assumption that the system is available for observation ahead of deployment.  
Thus, closed-loop control techniques are also application-specific and suffer from significant design-time overhead.

\subsection{Benefits of Model-independence}
A static model identified during development does not take into account complex system dynamics (e.g., variability between applications).
As the configuration space increases due to the increasing number of knobs, self-learning intelligent agents without apriori knowledge are attractive candidates to find optimal solutions through runtime observation. 
Reinforcement learning \cite{Sutton1998} is a prevalent candidate in the field of self-learning agents, demonstrating success for decision-making for services such as recommendation engines and games.
In this work, we utilize a model-free reinforcement learning approach to develop a model-free approximate memory controller that can learn the behavior of knobs through runtime experience. 
Model-independent control techniques can provide a general-purpose solution, independent of the application and system dynamics.

\section{Background and related work}
Approximate memory subsystems have been widely explored in the literature \cite{Mittal:2016:STA:2891449.2893356}. 
An integral step towards running an approximate application is to identify the non-critical sections of the data elements. 
Allowing faults in the critical data sections would lead to crashes and would require additional recovery mechanisms. Identifying non-critical data sections can either be done automatically \cite{Roy:2014:AAS:2597809.2597812, 10.1145/2463209.2488873, DBLP:conf/micro/VenkatagiriMHA16}, or through explicit programming language support  \cite{Ansel:2011:LCS:2190025.2190056}. 
Upon identifying the non-critical data, various memory approximation strategies can be implemented on systems, determined by underlying memory hardware technology. Table \ref{tab:tech} summarizes some of the standard technologies utilized throughout the memory hierarchy, along with the approximation objective. \seams{} is technology-agnostic and can leverage all of the technology knobs described in Table 1.

Several methods have been proposed in the context of tuning the memory approximation knobs. Table \ref{tab:comparison} identifies the most recent and relevant research, and compares \seams{} to these prior works. 
We define self-adaptivity as the ability to adapt to user-specified application goals or system constraints (e.g., increased target QoS).
We define self-optimization as the ability to find desirable system configurations given a fixed goal in the face of external disturbances (e.g., a scene change).
Manual schemes rely on designer expertise to optimize approximation knobs (e.g., $t_{REF}$ in DRAM). 
In EDEN \cite{10.1145/3352460.3358280}, Koppula \textit{et al.} show the effectiveness of manual tuning for neural networks, which have an intrinsic capacity of tolerating errors in memory accesses. 
EDEN uses approximate DRAM to reduce energy consumption and increase the performance of DNN inference. 
EDEN is limited to machine learning workloads and does not apply to a multi-level memory hierarchy. 
The absence of a runtime quality monitor in EDEN prevents dynamic reconfiguration of the approximation knobs (e.g., row activation delay $t_{RCD}$, operating voltage $V_{DD}$).

\begin{table}
  \caption{Memory approximation  approaches and the key challenges addressed ($\ast$ = uniquely addressed by \seams{}).}
  \label{tab:comparison}
    \centering
    \setlength\tabcolsep{2pt}
        \begin{tabular}{|l|c|c|c|c|c|}
            \hline
          &&&&&\\
            \textbf{Features} & 
            \textbf{EDEN} & 
            \stackanchor{\textbf{Control}}{\textbf{Theory}} & 
            \textbf{AdAM} & 
            \textbf{DART} & 
            \textbf{\seams{}}{} \\
            & \cite{10.1145/3352460.3358280} 
            & \cite{Maity2019} 
            & \cite{Teimoori2018}
            & \cite{Yarmand2020}
            &  \\
            \hline
            \hline
            \begin{tabular}{@{}l@{}}Technology Independent\end{tabular}
            &&\checkmark&&\checkmark&\checkmark \\\hline
            \begin{tabular}{@{}l@{}}Memory Hierarchy\end{tabular}
            &&&\checkmark&\checkmark&\checkmark \\\hline
            \begin{tabular}{@{}l@{}}Application Agnostic\end{tabular}
            &&&&&$\ast$ \\\hline
            Coordination &&&&&$\ast$ \\\hline
            Self-Adaptivity &&\checkmark&&&\checkmark \\\hline
            Self-Optimization &&&&&$\ast$ \\\hline
            \begin{tabular}{@{}l@{}}Model-Independence\end{tabular}
            &\checkmark&&&\checkmark&\checkmark \\\hline
            \begin{tabular}{@{}l@{}}Real System Evaluation\end{tabular}
            &\checkmark&&&&$\checkmark$ \\\hline
            
        \end{tabular}
\end{table}

Maity \textit{et al.} \cite{Maity2019} have proposed a solution to maintain a quality target at runtime by using classical control theory. 
Quality configuration tracking is modelled as a formal quality-control problem, and black-box modelling is used to capture memory approximation effects with variations in application input and system architecture. 
However, this scheme assumes only one level of the memory hierarchy is tuned at runtime and fails to address the problem of coordination between multiple knobs. 

In AdAM \cite{Teimoori2018}, Teimoori \textit{et al.}  
investigate memory approximation by managing approximation knobs across the memory hierarchy. 
AdAM solves a design-time ILP optimization problem and uses a runtime algorithm to adapt to new tasks by re-estimating the execution time. 
Although optimization techniques are a natural choice for simple architectural tuning, the lack of a feedback mechanism makes it too rigid for any sort of adaptivity (e.g., unknown inputs, and disturbances from other applications).
Furthermore, their  use-case only addresses a two-layer memory hierarchy, with an on-chip STT-RAM and an off-chip PCM Main Memory;
and the design-time algorithm is technology dependent. 

In DART \cite{Yarmand2020}, Yarmand \textit{et al.} propose a framework for a three-layer memory hierarchy (SRAM L1, SRAM L2, and an off-chip DRAM) without any technology-specific assumptions.
DART uses a branch and bound algorithm to consider all possibilities at design time, and creates a search tree to perform error probability analysis.
Although DART considers the full memory hierarchy, it requires the programmer to: (1) analyze the program during design time, (2) generate a memory profile for each application that would run on the system, and (3) estimate the worst-case probability of errors that would occur due to under-designed memory. 
Therefore DART requires apriori knowledge of the application and assumes that the system is available for full observation before deployment.

In the related topic of runtime resource management, machine learning approaches
have gained traction recently. 
Researchers have investigated the feasibility of machine learning methods for quality configuration in the approximation domain \cite{Masadeh2019, Masadeh2020}. 
However, conventional machine learning methods require extensive training to learn the correlation between the system's inputs and outputs. 
Static models that are defined ahead of deployment fail to handle new situations outside of expected behavior.
Online learning methods aim to address this issue and have shown promising results for resource management \cite{SOSA}. 

\seams{} incorporates the features highlighted in Table~\ref{tab:comparison} using online learning methods. 
The \seams{}' approach improves upon prior work by eliminating design-time modeling, being memory technology-agnostic, and coordinating multiple knobs at runtime to exploit approximation for multi-level memory hierarchies; enabling quick adoption of approximation for diverse platforms.
\section{\seams{} Methodology}
\label{sec:seams_methodology}

\begin{figure*}
    \centering
    \resizebox{\linewidth}{!}{
        \includegraphics[width=\linewidth]{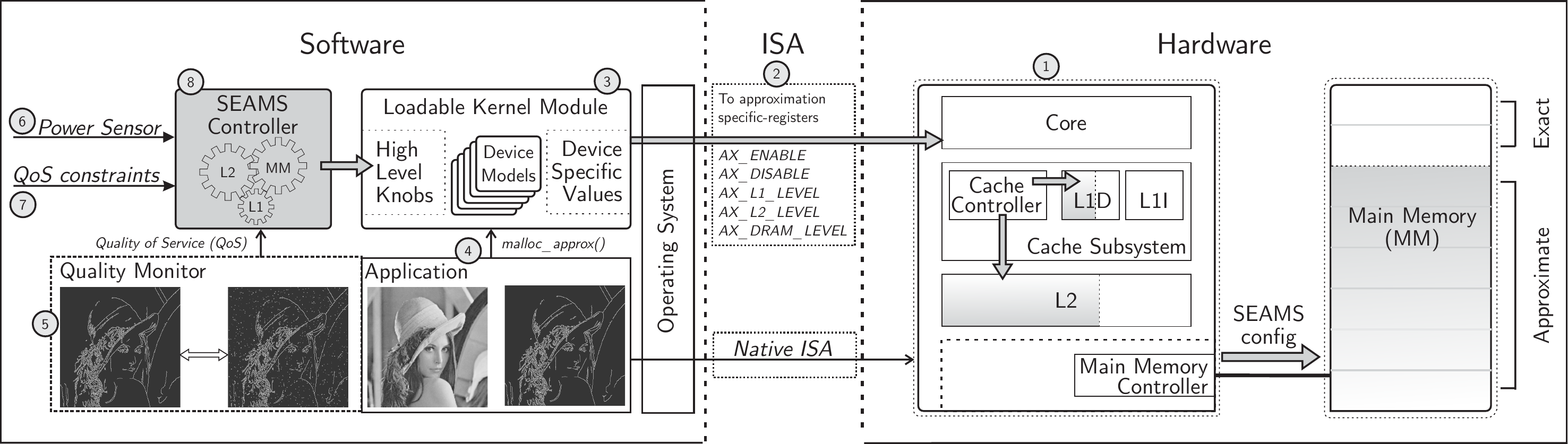}
    }
    \caption{Overview of \seams{} system architecture.
    }
    \label{fig:seams_system_arch}
\end{figure*}

Figure \ref{fig:seams_system_arch}  presents the 
\seams{}' realization of the logical  architecture described in Figure \ref{fig:single-knob}, consisting of the following components: 
\circled{1} a hardware platform with a processing unit, cache subsystem, and main memory. 
This hardware controls the degree of approximation at each memory layer by configuring the specific technology knobs available on the platform. 
Examples of technology knobs are in Table \ref{tab:tech}. 
The processor core contains special registers to set knobs (e.g., L1 $V_{DD}$ updated from \SI{0.7}{\volt} to \SI{0.8}{\volt}) through special instructions. 
For instance, in our current RISC-V 
realization of the processor, 
we deploy unused control and status registers (CSRs) for this purpose, as shown in Figure \ref{fig:seams_system_arch}. 
\circled{2} Instructions that write to these CSRs form an extension of the processor's ISA (RISC-V in our implementation), and are used to manage approximate elements at runtime. 
Truffle \cite{10.1145/2150976.2151008} is another example of a micro-architecture design that efficiently supports these ISA extensions for disciplined approximate programming. 
Instructions supported through these new CSRs include \texttt{AX\_ENABLE} to enable approximation, \texttt{AX\_DISABLE} to disable approximation,  \texttt{AX\_L1\_LEVEL} to set the technology-specific knob for Level 1 cache, \texttt{AX\_L2\_LEVEL} to set the knob for Level 2 cache, \texttt{AX\_DRAM\_LEVEL} to set the technology-specific knob for DRAM. 
\circled{3} A loadable kernel module that helps map the high-level knobs (e.g., low approximation) to technology-specific knob values (e.g., $V_{DD}$). 
For new technologies, the information in the module should be updated to reflect what the available actuation knobs are (e.g., available write pulse duration ($t_{write}$) for STT-RAM). 
The kernel module also allows applications to indicate which parts of the application's virtual memory can be placed physically in the approximate regions (explained further in Section \ref{sec:kernel_support}).
\circled{4} The user application running on this platform, specifying the non-critical sections of the data using an \texttt{malloc\_approx()} call to \circled{3} kernel module. 
\circled{5} A quality monitor computes the QoS periodically at runtime and reports it to \seams{}. 
\circled{6} The current power of the system is sensed using power sensors.
\circled{7} The expected QoS specific by the user. 
Expected QoS can be updated at runtime to adapt to different system objectives (e.g., a strict quality constraint optimizes \seams{} for more accurate executions, whereas a relaxed quality constraint optimizes \seams{} for energy savings).
\circled{8} The \seams{} Controller agent is the final component of the architecture and is responsible for runtime control of the memory approximation knobs.

The \seams{} controller agent is a model-free runtime manager for tuning configurable approximation knobs throughout the memory hierarchy. 
\seams{} follows the observe-decide-act (ODA) paradigm: the environment is observed through sensors during normal execution, and the decision-making agent is periodically invoked in order to (re)configure the system using knobs.
We design our decision-making logic by first defining our problem as a Markov Decision Process \cite{WHITE19891}: $(S,A,P_a,R_a)$, where $S$ refers to state space, $A$ refers to action space, $P_a$ refers to the transition probabilities from $S\rightarrow S'$ given action $A$, and $R_a$ refers to the expected rewards for selecting action $a$ in state $s$.
As is common when controlling real systems, we do not know the system dynamics and assume they change continuously.
This is a well-known problem, and to address it, we apply an appropriate established reinforcement learning solution, namely temporal difference (TD) learning \cite{10.1023/A:1022633531479}.

\subsection{Design Methodology}
Our goal is to design a decision-making agent that coordinates each layer in a unified 3-layer memory hierarchy to achieve acceptable application QoS while minimizing energy consumption.
First, we must define the structure of our environment.

\begin{figure}
    \centering
    \includegraphics[width=0.5\linewidth]{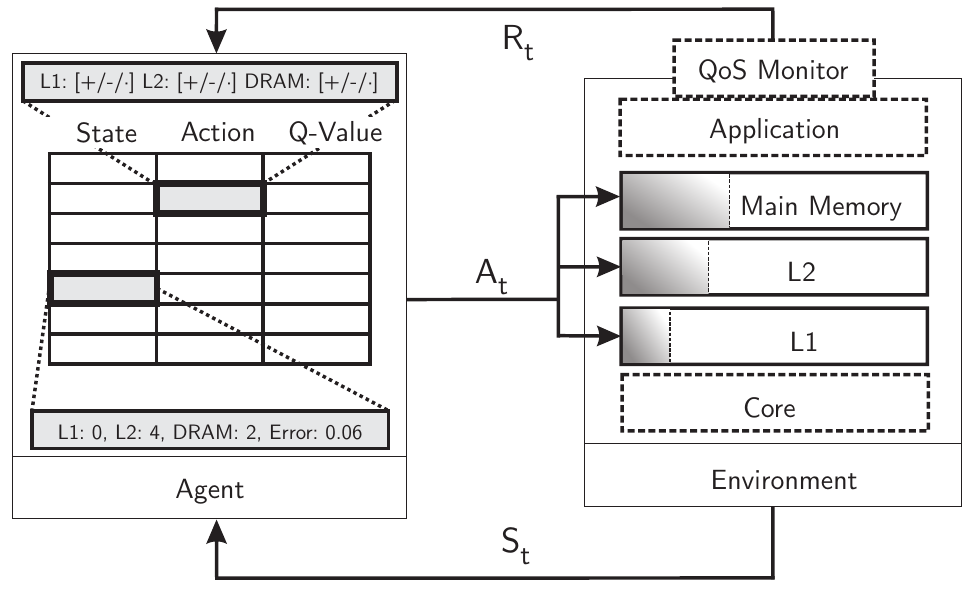}
    \caption{\seams{} taking actions against the environment, and the environment returns observations (updated state) and reward}
    \label{fig:seams_logic}
\end{figure}

\subsubsection{State Space ($S$)}
The state is a representation of the current system under control.
In \seams, we define a state vector that consists of high-level approximation settings (e.g., no/low/medium/high approximation) of each memory layer, as well as the current QoS error:
\begin{enumerate}
    \item L1D: current level 1 data cache configuration
    \item L2: current level 2 shared cache configuration
    \item Main memory: current main memory configuration
    \item Discretized QoS error ($Q_{threshold} - Q$), where $Q$ is the measured QoS, and $Q_{threshold}$ is the constraint) \end{enumerate}
This way, the state informs the agent what the current knob settings are, as well as how well they are achieving the goal of meeting the QoS requirement set by the application.
This allows us to translate the dynamics between application behavior and hardware configuration.
The QoS error is normalized to the worst-case QoS value ($max_Q$) to make \seams{} portable across applications, and high-level knobs allow \seams{} to be independent of memory technologies.

\subsubsection{Action Space ($A$)}
The action space contains all possible operations the agent may take to configure the system each time the agent is invoked.
The \seams{} action vector consists of the relative changes to the high-level knobs for layers in the memory hierarchy:
\begin{enumerate}
    \item L1D: Increase/Decrease/No change
    \item L2: Increase/Decrease/No change
    \item Main memory: Increase/Decrease/No change
\end{enumerate}
Initially, the \seams{} policy does not have any information regarding what actions are desirable and must discover which actions yield the maximum reward in each state via exploration (e.g., when there is no QoS constraint, actions which decrease power yield the maximum reward).

\subsubsection{Reward ($R$)}
\label{sec:method_reward}
The reward provides immediate feedback to the agent on how the previous state's action helped achieve the system goal.
In our case, this goal is to find the optimal configuration corresponding to minimum energy with acceptable QoS. 
Using the normalized power consumed measured at regular intervals, we define the reward in an unconstrained system by Equation~\ref{eqn:power_save}:
\begin{equation}
\label{eqn:power_save}
\begin{gathered}
    {reward}_P = 1 - \dfrac{Power}{max_{Power}} \\
    {reward}_P \in \{x| 0 \le x \le 1\} \\
\end{gathered}
\end{equation}
where $Power$ is measured power, $max_{Power}$ is the power consumed when the approximation is disabled at all layers of the memory hierarchy, and ${reward}_P$ is the reward obtained in terms of optimizing power.
This function represents a power optimization objective with a target power of zero. 
In an unconstrained system, operating at the highest power yields no reward, while operating at zero power yields the maximum reward. 
However, we must constrain the total reward in order to account for the quality threshold.

The policy should take actions which minimize the number of violations of the quality constraint specified by the application developer. 
Thus, the reward of a quality violation is calculated in Equation~\ref{eqn:quality_state} as:
\begin{equation}
\label{eqn:quality_state}
\begin{gathered}
    {reward}_Q = - \dfrac{Q - Q_{threshold}}{max_Q} \\
    {reward}_Q \in \{x| -1 \le x \le 1\}
\end{gathered}
\end{equation}
where ${reward}_Q$ is the reward obtained by staying within the quality constraint. 
In case of violations, ${reward}_Q$ is negative indicating that an undesired action was performed by \seams{}, which led to a QoS violation.

Finally, the reward (R) is calculated from the ${reward}_P$ and ${reward}_Q$ and reported to the agent by Equation~\ref{eqn:reward}:
\begin{equation}
\label{eqn:reward}
\begin{gathered}
 R= 
\begin{cases}
    {reward}_P,& \text{if } Q\le Q_{threshold}\\
    {reward}_Q,              & \text{otherwise}
\end{cases}
\end{gathered}
\end{equation}
\begin{algorithm}
	\caption{TD($\lambda$) algorithm \cite{Sutton1998} for determining \seams{} policy.}
	\label{algo:tdlambda}
	\begin{algorithmic}[1]
	    \State Algorithm parameters: step size, discount factor, trace decay $\alpha,\gamma,\lambda \in (0, 1]$
	    \State Initialize $Q(s,a)$ arbitrarily, for all $s \in \mathcal{S}, a \in \mathcal{A}(s)$
		\For {each episode}
		\State $E(s,a) =0, \forall s \in \mathcal{S}, a \in \mathcal{A}(s)$
		\State Initialize $S,A$
		\For {each step of episode}
		\State Take action $A$, observe $R$, $S'$
		\State Choose $A'$ from $S'$ using policy derived from $Q$
		\State $\delta \leftarrow R + \gamma Q(S',A') - Q(S,A)$
		\State $E(S,A) \leftarrow E(S,A) + 1$
		\For {each $s \in \mathcal{S}, a \in \mathcal{A}(s)$}
		\State $Q(s,a) \leftarrow Q(s,a) + \alpha \delta E(s,a)$
		\State $E(s,a) \leftarrow \gamma \lambda E(s,a)$
		\EndFor
		\State $S \leftarrow S';\; A \leftarrow A'$ 
		\EndFor
		\EndFor
	\end{algorithmic} 
\end{algorithm} 
\subsection{\seams{} Agent: Model-free Control}
\label{sec:modelfreecontrol}
Given the definition of the environment and goals, we simply need a decision-making mechanism (\seams) to find the optimal policy.
Initially, the \seams{} agent does not have any information regarding the environment and explores the state-space by taking purely arbitrary decisions (actions). 
It uses \textit{temporal-difference} (TD) learning \cite{Sutton1998} to learn directly from raw experiences without a model of the environment's dynamics. 
Figure~\ref{fig:seams_logic} shows the logical structure of the \seams{} agent and its relation to the environment, i.e., system under control. 
The agent interacts with the environment through actions, and the environment provides rewards and updated state information to the agent.

Actions that lead the system to optimize power without violating quality constraints are rewarded well.
The policy is modeled as a state-action value function by keeping track of all the state variables, along with the possible actions in the form of a table. 
Q-learning \cite{Watkins92q-learning}
is a popular TD control algorithm.
Q-learning aims to learn a state-action value function, $Q$, which directly approximates $q*$, the optimal state-action value function. 
A variation of Q-learning combines eligibility traces to obtain a more general method that may learn more efficiently. 
Eligibility traces look backward to recently visited states and act as short-term memory. 
This algorithm, where Q-learning is combined with a backward short-term memory using eligibility traces, is known as TD($\lambda$) \cite{10.1023/A:1022633531479}. 

\seams{} uses the TD($\lambda$) algorithm to update and optimize the approximation management policy throughout runtime continuously. 
The detailed algorithm is outlined in Algorithm \ref{algo:tdlambda}.
The dilemma presented during any controller designer is determining control parameters, whether the implementation uses classical control theory or reinforcement learning. 
In the TD($\lambda$) algorithm, learning parameters have interpretable meaning, so can be set several ways, e.g., using designer intuition or empirical observation.
In our case we determine learning parameters ($\alpha$=0.6, $\gamma$=0.1, and $\lambda$=0.95) empirically by simulating our control logic on system traces for \texttt{canny}. 
No matter the controller deployed, these parameters must be determined.
However, we define our control logic in such a way that the parameters apply to the \textit{type} of control (i.e., memory approximation knobs), as opposed to the \textit{application} under control (i.e., edge detection).

A Q-table is formed that maintains the Q-value of each state-action pair (Figure~\ref{fig:seams_logic}).
The agent is invoked periodically and performs the following steps during each invocation:
\begin{enumerate}
    \item Measure the power and QoS to evaluate the reward $R$
    \item Update the table ($Q$ values) based on reward $R$
    \item Sense the current approximation levels and QoS to determine the current state $S$
    \item Given the current state $S$ and updated $Q$ values, select next action $A$
\end{enumerate}

\begin{figure}
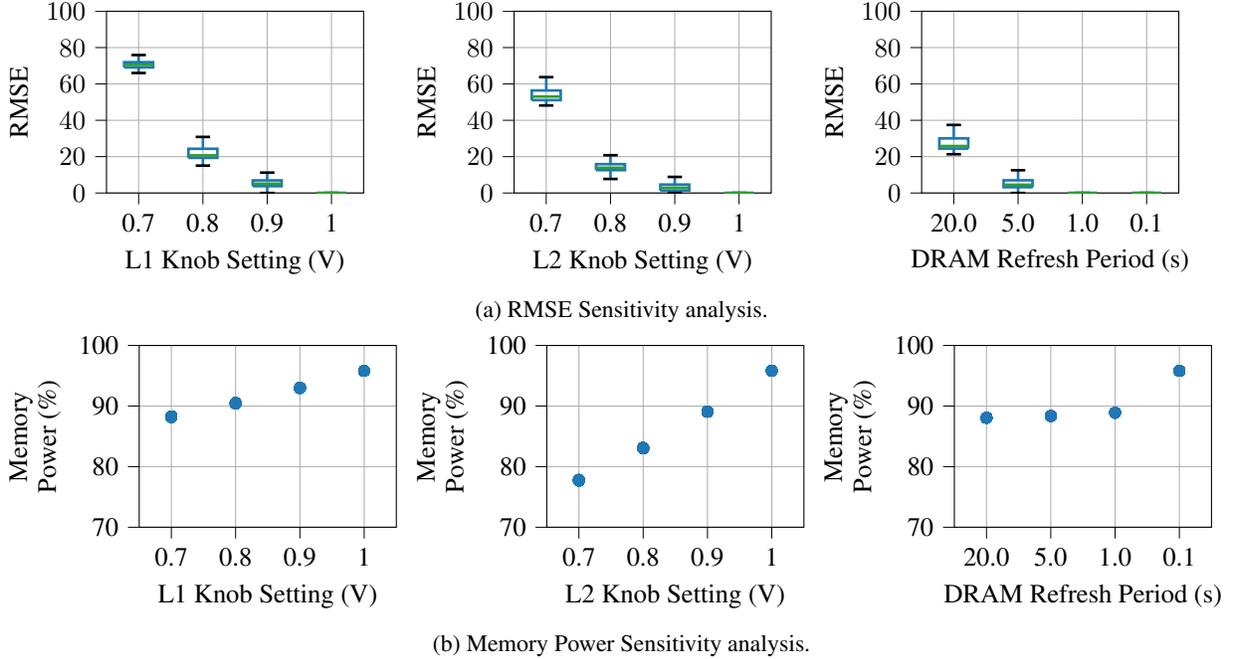

    \begin{subfigure}{\linewidth}
        \input{figs/sensitivity/rmse.tex}
        \caption{RMSE Sensitivity analysis.}
        \label{subfig:senseL1}
    \end{subfigure}
    \begin{subfigure}{\linewidth}
        \input{figs/sensitivity/power.tex}
        \caption{Memory Power Sensitivity analysis.}
        \label{subfig:senseDram}
    \end{subfigure}

    \caption{Sensitivity analysis of memory configuration knobs on QoS (RMSE, top) and memory power (normalized to {L1:\SI{1}{\volt}, L2:\SI{1}{\volt}, DRAM:\SI{0.064}{\second}}) for \texttt{canny}.}
    \label{fig:sensitivity}
\end{figure}

\subsection{\seams{} Case Study}
As described earlier, Figure \ref{fig:seams_system_arch}
outlines the \seams{} system architecture, with the \seams{} Controller agent
\circled{8} in software responsible for runtime control of the hardware memory approximation knobs
. Our implemented environment consists of a unicore RISC-V  processor with a three-layer memory hierarchy: L1 SRAM data cache, L2 SRAM shared cache, and DRAM main memory. \seams{} is implemented in software and runs in userspace. A Loadable Kernel Module accompanies \seams{} which incorporates the device specific translations from high-level configurations (e.g., \SI{25}{\%} approximation) to technology specific values (e.g., \SI{0.9}{\volt} for SRAM caches). 

The state vector $S$ is made up of (1) high-level configurations corresponding to memory layers, and (2) the QoS error.
Discrete integer values represent all of the vector values.
The L1 and L2 voltage levels ($V_{DD}$) are between 0.7-\SI{1.0}{\volt} in increments of \SI{0.1}{\volt} \cite{Yarmand2020}.
The main memory refresh periods are \SI{0.1}{\second}, \SI{1}{\second}, \SI{5}{\second}, \SI{20}{\second} \cite{Liu2011}.
The QoS error is normalized and discretized into 16 buckets of step size $log_{2}16$.
The inclusion of the QoS error in the state differentiates desirable actions for the same voltage level, depending on the QoS error explicitly.

The action vector $A$ contains a field for adjusting each of the L1, L2, and main memory knobs.
The possible knob configurations are voltage levels for cache L1 and L2 and refresh periods for DRAM main memory.
Actions for each knob only consist of increase by one, decrease by one, or remain the same.
To keep the action space manageable, we performed a sensitivity analysis on the knob in each memory layer.
Figure \ref{fig:sensitivity} shows the sensitivity analysis of the approximation knobs on the output QoS and power.
We make three observations.
(1) As we move up the memory hierarchy (i.e., from L1 to L2 to main memory), the quality is less affected by higher degrees of approximation. 
(2) The contribution of memory power from individual levels of hierarchy varies. 
Although main memory techniques can save around 23\% DRAM power, when the full memory hierarchy is considered, DRAM's power savings saturates at 12\%.
(3) Having four knob configurations captures the range of power/quality trade-off effects while keeping the state-space manageable.
We conclude that four knob configurations for each level provide sufficient control for reaching our goal.

Reward $R$ is calculated based on Equation~\ref{eqn:reward}.
To evaluate the reward, \seams{} uses software level sensors to determine the application's output quality. 
Although the metric is domain-specific and is generated by a quality monitor, normalizing it to the worst quality keeps \seams{} domain agnostic. 
We update $Q$ values using the reward as specified in Algorithm~\ref{algo:tdlambda}.

To demonstrate the efficacy of \seams{} for coordinating knobs in the memory hierarchy, we deploy a hardware platform that mimics the effects of approximation. 
The effect of approximation knobs in each layer in the memory hierarchy is determined using existing models in literature \cite{Yarmand2020,Liu2011}. 

\section{Experimental setup}
\label{experimentalSetup}
For the \seams{} architecture described in Section \ref{sec:seams_methodology},
we describe our experimental setup for the RISC-V hardware platform on which \seams{} is running (\circled{1} in Figure \ref{fig:seams_system_arch}). 
An overview of our evaluation platform is shown in 
Figure \ref{fig:ariane-mod}. 
We implement \seams{} in software running on Openpiton \cite{Balkind}, an open-source framework designed to enable scalable architecture research prototypes. 
We use Openpiton with a single Ariane \cite{Zaruba2019} core, a 64-Bit RISC-V core capable of running Linux. 
The framework is synthesized on a DIGILENT NexysVideo board, having a Xilinx Artix-7 FPGA(XC7A200T-1SBG484C). 
The parameters used to synthesize the system are summarized in Table~\ref{tab:config}.

\begin{scriptsize}
\begin{table}[tb]
\caption{System configuration used for \seams{} evaluation.}
\label{tab:config}
\centering
    \begin{tabular}{cl}
    \toprule
    Component & Configuration \\
    \midrule
    Cores & 1 \\
    TLBs & Number of entries (16) \\
    L1 D-Cache  & Number of sets, ways (\textbf{16kB, 4-way}) \\ 
    L2 Cache  & Number of sets, ways (\textbf{64kB, 4-way}) \\ 
    Floating-Point Unit & Present \\ 
    Main Memory & Onboard (512MB 800MHz DDR3) \\ 
    Clock frequency & 30 MHz \\ 
    \bottomrule
    \end{tabular}
\end{table}
\end{scriptsize}

\subsection{Modifications to RISC-V Core}
We further modify the Ariane core to support fault injection throughout the memory hierarchy. The synthesized core, running on the NexysVideo board, does not have an option to configure real knobs for approximation. However, we rely on existing works that map device-specific approximation knobs to the observable bit error rates \cite{Yarmand2020,Liu2011}. 
Thus, we introduce bit errors through fault injection in order to emulate the effect of approximation knobs.

\subsubsection{SRAM Fault Injection}
\begin{figure}[!tb]
    \centering
    \resizebox{0.5\linewidth}{!}{
    \includegraphics[width=\linewidth]{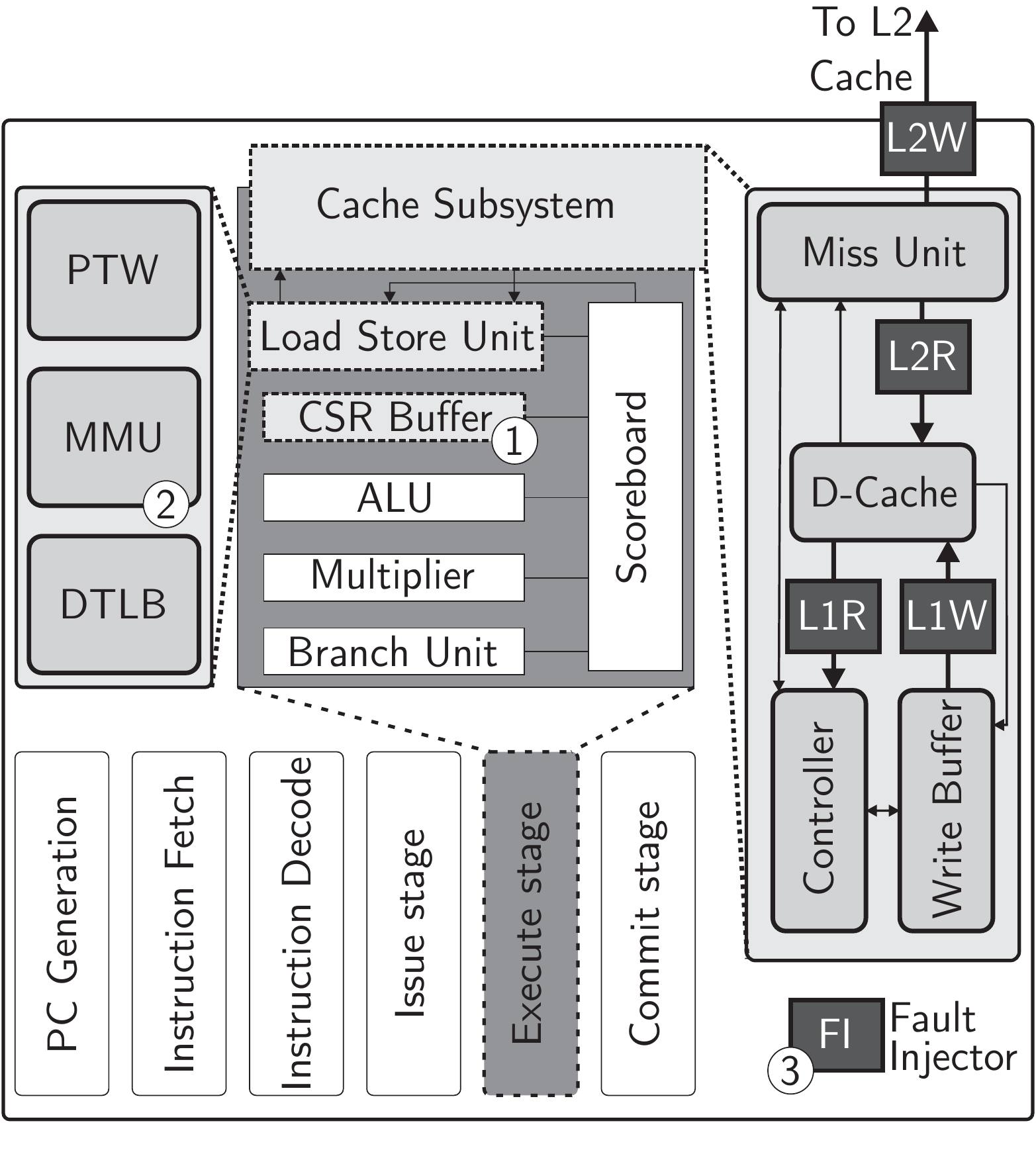}
    }
    \caption{Modification of Ariane RISC-V core to emulate on-chip approximate memory. \circled{1} Addition of new CSRs to communicate with \seams{} kernel module. \circled{2} Modification of address translation logic in Memory Management Unit (MMU) to generate \texttt{approx} signal. \circled{3} Fault injectors which introduce errors in the memory bus.}
    \label{fig:ariane-mod}
\end{figure}

The RISC-V specification defines separate addresses for Control and Status Registers (CSRs) associated with each hardware thread \cite{Waterman2019}. 
Unused CSRs are utilized by the kernel to communicate information required for the configuration of the approximation knobs. In Figure~\ref{fig:ariane-mod} additional CSRs are denoted with \circled{1}.
In particular, the following information is stored in CSRs:
\begin{enumerate}
    \item L1 data cache Read and Write Bit Error Rate
    \item L2 shared cache Read and Write Bit Error Rate
    \item Starting and Ending physical address of the non-critical memory segment
\end{enumerate}

The Bit Error Rates correspond to specific memory nodes and are translated from technology-specific values described in Section \ref{sec:error_model}. The information from the CSRs is propagated to the \circled{2} Memory Management Unit (MMU), where address translation takes place. 
MMU uses this information to generate an additional \texttt{approx} bit along with the \texttt{index} and \texttt{tag} bits to indicate that this address in the valid range of approximation. 
The \texttt{approx} bit generation is repeated whenever a virtual address is converted to a physical address. 
The \texttt{approx} bit in conjunction with the CSRs for Bit Error Rate is utilized by the cache controller to control the degree of approximation and contain it to the non-critical parts of the application. 
A Fault Injector (\texttt{FI}) module is used to emulate the effects of approximation by introducing the bit flips on the memory bus. 
Four \texttt{FI} modules are instantiated in the cache subsystem as shown in Figure~\ref{fig:ariane-mod} \circled{3}. 
The \texttt{FI} modules generate a bit-flip mask for each memory access using a Linear-Feedback Shift register (LFSR)
that introduces randomness in the injected errors. 

The \texttt{FI}s are located in (1) Data Cache Memory emulating the bit flips corresponding to L1 data reads and L2 reads, (2) Write Buffer emulating the bit flips corresponding to L1 data writes, and (3) Miss Unit emulating the bit flips corresponding to L2 writes.

\subsubsection{DRAM Fault Injection}
DRAM cells store data in capacitors that lose charge over time. 
In order to keep the data consistent, the DRAM cells have to be refreshed periodically. 
DRAM cells' strength is non-uniform due to manufacturing variability, i.e., some DRAM cells lose charge faster than others. 
The number of bit-flips in DRAM increases as the refresh period increases due to a higher number of DRAM cells losing charge before they are refreshed. 
These bit-flips also depend on when the data was written into and read from the DRAM. 
Therefore, implementing a \texttt{FI} module for DRAM requires keeping track of the faulty DRAM cells for each refresh-rate knob and the hold times of the data in each DRAM cell. 
Given the DRAM size, maintaining this information requires a lookup table of impractical size on an FPGA.
The lookup table also introduces a considerable latency in DRAM reads/writes. 
To emulate DRAM errors, we implement a software-based \texttt{FI} for DRAM. 
Initially, a map \texttt{DRAM\_MAP} of faulty DRAM cells for the maximum refresh period (\SI{20}{\second}) knob is generated randomly using a uniform distribution. 
The faulty DRAM cells for higher refresh rates are a subset of \texttt{DRAM\_MAP}. The data being loaded into the DRAM is modified using the \texttt{DRAM\_MAP} and the current refresh rate knob. 
The exact read and write accesses to the DRAM are not impacted.

\begin{scriptsize}
    \begin{table}[tb]
    \caption{Applications used for \seams{}' evaluation along with their inputs and QoS.}
    \label{tab:apps}
    \centering
        \begin{tabular}{llll}
        \toprule
        Application & Domain & Input Size & Quality Metric \\
        \midrule
        canny\cite{Canny} & Image-Processing & 352x288 (Grayscale) & Image Diff (RMSE) \\
        k-means\cite{7755728} & Machine Learning & 426x240 (RGB) & Image Diff (RMSE) \\
        blackscholes \cite{7755728}  & Finance  & 4K entries & Avg. Relative Error \\ 
        \bottomrule
        \end{tabular}
    \end{table}
\end{scriptsize}

\subsection{User application and QoS metric}
\label{experimentalSetup_app}
The \seams{} methodology is well suited for a large class of workloads that have a high intensity of memory operations (e.g., video processing, machine learning).
Table \ref{tab:apps} summarizes the applications used for \seams{}' case study. (1) \texttt{Canny} edge detection \cite{Canny} algorithm operates on a video streams and marks the edges in each frame. 
(2) \texttt{k-means} is a machine-learning application \cite{7755728} which partitions 3 dimensional input points (RGB pixels) into 6 different clusters, and (3) \texttt{blackscholes} \cite{7755728} is a finance analysis application which solves partial differential equations to perform price estimations. 

The applications' source code is modified to indicate which data elements are non-critical. 
Several techniques have been explored in the literature \cite{Venkatagiri2019, 8464151} 
to systematically analyze and report how different parts of the application are affected by errors. 
Depending on the application, there are one or more candidate data segments (e.g., image data, video data, signal data) for accuracy/energy tradeoffs. 
We identify these segments in the source code, and replace \texttt{malloc()} calls to the kernel by \texttt{malloc\_approx()} calls. 
For \texttt{canny}, the image buffer is marked as a non-critical section.
For \texttt{k-means}, the data structure for the image buffer is modified to separate the non-critical pixel data, and raw pixel information is converted from float representation to unsigned char representation since each pixel value lies between 0 and 255. For \texttt{blackscholes}, the buffer data structure is left unmodified: the non-critical approximate memory consists of a buffer of floats. Thus, errors can impact the bits differently, and in case of extreme approximation may produce relative errors of $100\%$.
The \texttt{malloc\_approx()} calls are intercepted by a custom Linux Kernel Module, described in next section.

In addition to specifying the non-critical data elements, a quality-monitor specific to the application domain is required. 
The quality monitor is a lightweight software routine invoked to evaluate the application QoS and used to calculate the reward as described in Section \ref{sec:method_reward}. 
The QoS metric indicates the quality degradation caused by the configuration of approximation knobs. 
Typically, application developers generate a software routine that is capable of measuring the quality at runtime. In \texttt{canny}, the QoS is determined by evaluating the \textit{Root Mean Square Error} (RMSE), which is the mean of pixel differences squared between an exact result and an approximate result.
For \texttt{k-means} and \texttt{blackscholes} the quality monitors are RMSE and \textit{Average Relative Error}, used directly from AxBench \cite{7755728}.
This software routine is invoked during runtime, and the result of an exact run of the application is compared with an approximate version \cite{Green-Baek-PLDI-2010, 10.1007/978-3-319-52709-3_7}. 
The quality evaluation is not repeated for every input so that the benefits of approximation can justify the overhead. 
Depending on the status of learning, the frequency of quality evaluation should be adapted. A detailed overhead analysis is presented in Section \ref{sec:overhead}.
If additional cores are available, ground truth comparison can be performed in parallel.
In our unicore setup, this incurs unavoidable overhead during the initial exploration phase.

\subsection{Kernel Support for Approximation Knobs}
\label{sec:kernel_support}
We develop a Loadable Kernel Module (LKM) as middleware between the user application and CSRs. \texttt{malloc\_approx()} calls from the applications are intercepted by the LKM, and using \texttt{mmap} a contiguous physical segment is allocated. 
The starting and ending address of the segment is written to additional CSRs.
Whenever the user application loads/stores data, the MMU compares the memory address in hardware using these CSRs to check if it is in the non-critical segment. 

\subsection{Power Model}
\label{sec:power_model}
The evaluation platform does not come equipped with on-board power sensors. 
Thus, we use Sniper \cite{carlson2014aeohmcm} simulations and McPAT \cite{10.1145/1669112.1669172}, along with existing power models from literature \cite{Yarmand2020,Liu2011} 
for computing the power for different knob settings. Each new input to the application is simulated in Sniper, and McPAT is invoked to estimate the power and energy consumption of different system components. The power and energy are then scaled according to the technology models (described below) to estimate different knob configurations' power values.

\begin{figure*}
    \begin{subfigure}[b!]{0.45\linewidth}
        \begin{tikzpicture}
    \begin{semilogyaxis}[grid,
            height = 4.5cm,
            xlabel=Relative Power Supply Voltage ($\%$),
            ylabel style={align=center}, ylabel=SRAM\\Bit Error Rate,
            xmin=40,
            xmax=100,
            xtick = {40,50,60,70,80,90,100},
            ytick = {1e-7,1e-6,1e-5,1e-4,1e-3,1e-2,1e-1,1e0},
            legend style={fill opacity=0.8, draw opacity=1, text opacity=1, draw=white!80!black, font=\small},
        ]

        \addplot[smooth, color=blue,mark=square*] coordinates {
                (50, 3.55E-01)
                (60, 8.71E-02)
                (70, 7.59E-03)
                (80, 3.24E-04)
                (90, 2.00E-05)
                (100,2.00E-06)
            };
        \addlegendentry{Read Error}

        \addplot[smooth, color=red,mark=*] coordinates {
                (50, 3.33E-02)
                (60, 1.00E-03)
                (70, 1.00E-04)
                (80, 2.00E-05)
                (90, 1.00E-06)
                (100,0)
            };
        \addlegendentry{Write Errors}

    \end{semilogyaxis}
\end{tikzpicture}
        \caption{Bit error rate for a 6T SRAM cell with varying $V_{DD}$ values in 65nm. Data from \cite{Yarmand2020}. }
        \label{fig:sramModel}
    \end{subfigure}
    \hspace{0.5cm}
    \begin{subfigure}[b!]{0.45\linewidth}
        \begin{tikzpicture}
    \begin{axis}[grid,
            height = 4.5cm,
            xlabel=Refresh Cycle (s),
            xtick={1,2,3,4,5,6,7,8},
            xticklabels={0.1,0.2,0.5,1,2,5,10,20},
            ylabel style={align=center}, ylabel={Self-refresh\\Power Saving (\%)},
            ymin=0,
            ymax=30,
            ytick={0,5,10,15,20,25,30},
            yticklabels={0,5,10,15,20,25,30},
        ]

        \addplot[smooth, color=blue,mark=triangle*] coordinates {
                (1, 8.42007434944238)
                (2, 16.003717472118957)
                (3, 20.687732342007436)
                (4, 22.323420074349443)  
                (5, 23.066914498141266)
                (6, 23.438661710037174) 
                (7, 23.66171003717472)
                (8, 24)                 
            }; 
            \label{plot_power}
         	\addlegendentry{Power saving}
    \end{axis}

    \begin{semilogyaxis}[
            height = 4.5cm,
            axis x line=none,
            ylabel=Error Rate,
            ylabel near ticks,
            yticklabel pos=right,
            ytickten={-11,-9,-7,-5,-3,-1},
            ymax=1,
            legend style={at={(0.98,0)},anchor=south east, fill opacity=0.8, draw opacity=1, text opacity=1, draw=white!80!black, font=\small}
        ]
        \addplot[smooth, color=red,mark=square*, mark options={fill=white}] coordinates {
                (1, 8.088161777987015e-11)
                (2, 5.007935227340456e-10)
                (3, 7.96203148797984e-9)
                (4, 3.8336966683708786e-8)
                (5, 2.527728210797576e-7)
                (6, 0.0000037739123458197472)
                (7, 0.000020605975259769508)
                (8, 0.00014468035186892834)
            };
        \label{plot_error}
        \addlegendentry{Error Rate}
    \end{semilogyaxis}

\end{tikzpicture}
        \caption{Bit error rate and power savings for different refresh cycles in DRAM array. Data from Flikker \cite{Liu2011}. }
        \label{fig:dramModel}
    \end{subfigure}
\end{figure*}

\subsection{Error Model}
\label{sec:error_model}
\textit{On-chip SRAM}: When scaling the supply voltage ($V_{DD}$) in SRAM cells, read and write errors are dominant; hence hold failures are not considered here. 
We use a model for a 6T SRAM for \SI{65}{\nano\meter} node from \cite{Yarmand2020} for comparison with related memory approximation work. 
The Bit Error Rate corresponding to relative power supply voltage is shown in Figure \ref{fig:sramModel}.

\textit{Off-chip DRAM}: We employ the power model proposed in Flikker \cite{Liu2011}. 
We assume that the DRAM is partitioned into two sections: (1) 1/4 exact DRAM having a high refresh rate, and (2) 3/4 DRAM having a lower refresh rate based on the approximation knob. 
The corresponding power model is shown in Figure \ref{fig:dramModel}.

\section{Performance Evaluation}
In this section, we demonstrate \seams{}’ ability to learn directly from raw experience, without requiring any model of the environment's dynamics. These experiments are evaluated against \texttt{canny}.

\subsection{Policy Initialization}
\label{sec:exp:phase0}
First, we evaluate \seams{} ability to learn an optimal policy to minimize energy from scratch.
We compare two TD reinforcement learning algorithms: TD($\lambda$) and Q-Learning.
The primary difference between the methods is that TD($\lambda$) uses bootstrapping.
For both algorithms, we determined the learning parameters empirically using a simulated workload (discussed in Section \ref{sec:modelfreecontrol}).
\begin{figure}[!bt]
\begin{subfigure}[t]{\linewidth}
    \centering
    \resizebox{\linewidth}{!}{
\begin{tikzpicture}

\definecolor{color0}{rgb}{0.12156862745098,0.466666666666667,0.705882352941177}
\definecolor{color1}{rgb}{1,0.498039215686275,0.0549019607843137}

\begin{axis}[
height=4cm,
legend cell align={left},
legend style={fill opacity=0.8, draw opacity=1, text opacity=1, draw=white!80!black},
minor xtick={},
minor ytick={},
tick align=outside,
tick pos=left,
width=\linewidth,
x grid style={white!69.0196078431373!black},
xlabel={Frame \#  (divided by 100)},
xmajorgrids,
xmin=-4.95, xmax=103.95,
xtick style={color=black},
xtick={-20,0,20,40,60,80,100,120},
y grid style={white!69.0196078431373!black},
ylabel style={align=center}, ylabel={Memory Power (\%)},
ymajorgrids,
ymin=0.55, ymax=0.85,
ytick style={color=black},
ytick={0.55,0.60,0.65,0.7,0.75,0.8,0.85},
yticklabels={55,60,65,70,75,80,85},
]
\addplot [line width=1pt, color0]
table {%
0 0.758095734834559
1 0.804050742187501
2 0.782084327205878
3 0.748171426011027
4 0.761010072610293
5 0.73763439751838
6 0.720274765624998
7 0.738340546875001
8 0.710776396599267
9 0.712783958180143
10 0.732561877757356
11 0.716128141084558
12 0.696869172334556
13 0.701672051470585
14 0.695716382352939
15 0.690132459558824
16 0.679105511948528
17 0.683335106617643
18 0.670170820312502
19 0.688517531709559
20 0.674080234374999
21 0.65597536948529
22 0.661234621323525
23 0.669282683823528
24 0.670473822610292
25 0.66456005330882
26 0.666975300551472
27 0.659350820772058
28 0.662422082720586
29 0.652066517463232
30 0.651471410845585
31 0.652259768382351
32 0.655503959099263
33 0.65403517922794
34 0.656957676470587
35 0.661147000919116
36 0.657724769761027
37 0.657214177849262
38 0.655634168198526
39 0.651322267922792
40 0.65093619852941
41 0.655466086397057
42 0.647226252757351
43 0.646156377757351
44 0.642751838235293
45 0.644388097426469
46 0.650342340073529
47 0.64796799632353
48 0.647593300551473
49 0.647591083639707
50 0.639614471507352
51 0.642466451286765
52 0.645125459558823
53 0.641131065257352
54 0.642502524816175
55 0.64187012867647
56 0.641963556985293
57 0.64242150735294
58 0.644274862132352
59 0.644335156249999
60 0.643833118566175
61 0.640624218749999
62 0.640512545955881
63 0.64147757352941
64 0.645384283088234
65 0.64712306985294
66 0.645877481617647
67 0.644050672794118
68 0.641111672794117
69 0.643212637867647
70 0.639951792279411
71 0.643286121323529
72 0.647099724264707
73 0.645722056985296
74 0.644374446691178
75 0.63984825367647
76 0.640896783088234
77 0.641591156249999
78 0.63978478860294
79 0.638272517463234
80 0.641694147058823
81 0.638514950367645
82 0.639516648897058
83 0.639549462316175
84 0.638088295955882
85 0.640605252757352
86 0.639865670036763
87 0.638580023897058
88 0.640363801470586
89 0.63955585110294
90 0.641161899816175
91 0.639616480698528
92 0.638164123161764
93 0.638985294117646
94 0.638807275735293
95 0.637261365808823
96 0.637843396139705
97 0.637516512867646
98 0.63857934742647
99 0.64508262867647
};
\addlegendentry{Q-Learning}
\addplot [line width=1pt, color1]
table {%
0 0.767660702665439
1 0.763047387867646
2 0.74431425229779
3 0.745549832261029
4 0.730878903952205
5 0.735573764246325
6 0.717811421875
7 0.718758315716912
8 0.703437937499998
9 0.701927324448529
10 0.693204996783089
11 0.690906337316178
12 0.688709303308823
13 0.692287727941179
14 0.686847038602939
15 0.67939264016544
16 0.694662253216915
17 0.676206772518382
18 0.673543637867645
19 0.673293201746321
20 0.667287137408087
21 0.674771180147057
22 0.667420580882353
23 0.677206094209559
24 0.672555678768382
25 0.666684852481617
26 0.658834808363969
27 0.66462583180147
28 0.670881738970588
29 0.660700386029407
30 0.658130682904408
31 0.662385338235293
32 0.658223178768378
33 0.656889056525734
34 0.651862152573525
35 0.660274910386028
36 0.665000723345587
37 0.662168216911763
38 0.654459969669115
39 0.654266946691174
40 0.650317046415438
41 0.65383403400735
42 0.664150443014705
43 0.654366039522056
44 0.651366913602939
45 0.646592113970585
46 0.651307900275733
47 0.649395604779411
48 0.645720375919116
49 0.644936377757353
50 0.647846999080879
51 0.642772701286764
52 0.644165761948528
53 0.643713309742645
54 0.646894165441175
55 0.641672068933823
56 0.644042552389705
57 0.64246085110294
58 0.645562738511029
59 0.643475873161763
60 0.643250505514705
61 0.641237362132352
62 0.641113143382352
63 0.648907031249999
64 0.640583547794118
65 0.645662178308822
66 0.643109145220587
67 0.639049094669117
68 0.640963465073528
69 0.64041135110294
70 0.638536779411764
71 0.637337086397058
72 0.638935110294115
73 0.64240842463235
74 0.643103660845585
75 0.640810156249997
76 0.639548759191174
77 0.640348207720587
78 0.641551424632351
79 0.642557766544117
80 0.642142463235292
81 0.638611503676469
82 0.64117616360294
83 0.639353783088234
84 0.637483593749999
85 0.638814200367646
86 0.638497564338235
87 0.639293795955882
88 0.640511855698529
89 0.639013832720588
90 0.638847951286764
91 0.640802537224264
92 0.636532642463235
93 0.638307612132352
94 0.637313373161764
95 0.637379779411765
96 0.637433945772059
97 0.636557628676471
98 0.645077780330882
99 0.638719669117646
};
\addlegendentry{TD($\lambda$)}
\addplot [line width=1pt, black, dashed]
table {%
0 0.625
100 0.625
};
\addlegendentry{Optimal}
\end{axis}

\end{tikzpicture}
    }
    \caption{Configuration space = 64 states}
    \label{fig:exp:0:subfig:small}
\end{subfigure}%

\begin{subfigure}[t]{\linewidth}
\centering
    \resizebox{\linewidth}{!}{
\begin{tikzpicture}

\definecolor{color0}{rgb}{0.12156862745098,0.466666666666667,0.705882352941177}
\definecolor{color1}{rgb}{1,0.498039215686275,0.0549019607843137}

\begin{axis}[
height=4cm,
legend cell align={left},
legend style={fill opacity=0.8, draw opacity=1, text opacity=1, draw=white!80!black},
minor xtick={},
minor ytick={},
tick align=outside,
tick pos=left,
width=\linewidth,
x grid style={white!69.0196078431373!black},
xlabel={Frame \#  (divided by 100)},
xmajorgrids,
xmin=-4.95, xmax=103.95,
xtick style={color=black},
xtick={-20,0,20,40,60,80,100,120},
y grid style={white!69.0196078431373!black},
ylabel style={align=center}, ylabel={Memory Power (\%)},
ymajorgrids,
ymin=0.55, ymax=0.85,
ytick style={color=black},
ytick={0.55,0.60,0.65,0.7,0.75,0.8,0.85},
yticklabels={55,60,65,70,75,80,85},
]

\addplot [line width=1pt, color0]
table {%
0 0.788994896599262
1 0.823323724724269
2 0.803139290900732
3 0.824188720588234
4 0.819786120404416
5 0.804483270680148
6 0.816034830882353
7 0.780537912683821
8 0.802914301930148
9 0.780245008731611
10 0.759549701746321
11 0.768022841452205
12 0.774844834558817
13 0.769682936580879
14 0.757298600643376
15 0.759528969209554
16 0.782477147518378
17 0.750612293658087
18 0.742027606158085
19 0.770133987591906
20 0.739624616268382
21 0.747580051930146
22 0.74393613419117
23 0.74872305055147
24 0.778066871783089
25 0.743882051470591
26 0.731191982996317
27 0.7468602265625
28 0.751912837775738
29 0.746757929687499
30 0.737476475183821
31 0.733181283088235
32 0.745729415900734
33 0.732915981158084
34 0.741276689338233
35 0.731529904871326
36 0.726912471966909
37 0.729023311580882
38 0.720031799632351
39 0.74267628446691
40 0.721885937499998
41 0.711343680147055
42 0.713869205422794
43 0.708859454963234
44 0.730383725643379
45 0.722990020680146
46 0.725604182904411
47 0.722353072150737
48 0.725172838694852
49 0.70166694806985
50 0.721174674632352
51 0.7048473359375
52 0.691984674632353
53 0.720843129595588
54 0.697977088694853
55 0.697729886948531
56 0.725317895680149
57 0.712822694852943
58 0.714298326286774
59 0.705875987132355
60 0.696528015624999
61 0.706538585018383
62 0.701675945772057
63 0.700758413602937
64 0.700490367187494
65 0.705028710477936
66 0.702332317555144
67 0.688865545496321
68 0.712427943474264
69 0.680505990808824
70 0.682820506893385
71 0.695568613970591
72 0.700439057904411
73 0.683700167738967
74 0.696143772058824
75 0.685022093290445
76 0.672653607996322
77 0.678151077665439
78 0.678497345128676
79 0.679719431985291
80 0.692207672334556
81 0.675705456341908
82 0.694802352481616
83 0.674580721966911
84 0.693178491727943
85 0.683398032628676
86 0.650086962316177
87 0.656926497242646
88 0.661273434283084
89 0.660145279871322
90 0.656960107077203
91 0.67684944898897
92 0.658408944393379
93 0.667738568474262
94 0.658117355698529
95 0.665611234834557
96 0.657063994485292
97 0.64846837040441
98 0.682416691636025
99 0.651933880055145
};
\addlegendentry{Q-Learning}
\addplot [line width=1pt, color1]
table {%
0 0.788466429227943
1 0.782908914062498
2 0.785842965533086
3 0.779441245863966
4 0.782009398897062
5 0.747834424632352
6 0.746338170496319
7 0.745888496323523
8 0.755299030330883
9 0.752890710477939
10 0.752084076286761
11 0.758977478860292
12 0.742621174172795
13 0.765478537683822
14 0.757669667738967
15 0.758740984374999
16 0.756554951746321
17 0.762732514705879
18 0.756492390624996
19 0.73954380836397
20 0.74240500321691
21 0.752887245404407
22 0.730598984834557
23 0.742356863511027
24 0.741356994485298
25 0.724385686121324
26 0.736797314338231
27 0.742204219209557
28 0.715800491268381
29 0.729311487132348
30 0.716491252297796
31 0.706381635569852
32 0.706336287224262
33 0.699706933363967
34 0.695259788602941
35 0.709126589154409
36 0.696150017003678
37 0.702568985294119
38 0.690538732077204
39 0.688311869944852
40 0.684886832720587
41 0.690905687040442
42 0.685482372242644
43 0.689706892003674
44 0.692242801470588
45 0.690520165441177
46 0.678883219209559
47 0.682547420036762
48 0.690218043658088
49 0.683309066176469
50 0.683149946231614
51 0.688610309283089
52 0.682724614430149
53 0.68190604733456
54 0.6770466484375
55 0.677654511488966
56 0.675149642463232
57 0.667998463694851
58 0.661353544117645
59 0.656909114430143
60 0.660806673253673
61 0.665434333639704
62 0.661072162683822
63 0.658443461856615
64 0.653845575367646
65 0.67211560799632
66 0.651734680606616
67 0.666912070312498
68 0.663481653952207
69 0.664807944393383
70 0.651473732536763
71 0.666608251378675
72 0.660757009191177
73 0.658053751838235
74 0.65718023897059
75 0.64809858639706
76 0.649041522977942
77 0.650957748621322
78 0.654234888786763
79 0.655087288602946
80 0.651587405330882
81 0.649762483455882
82 0.645777450367646
83 0.645768237132351
84 0.646707246323527
85 0.652142917279411
86 0.654508522058822
87 0.644903762867644
88 0.64951371415441
89 0.642129067095586
90 0.644108056066176
91 0.64920005009191
92 0.642561534466909
93 0.644730305147057
94 0.64558802665441
95 0.642295795036766
96 0.649762588235292
97 0.648250330422795
98 0.643314363051469
99 0.647870094669116
};
\addlegendentry{TD($\lambda$)}
\addplot [line width=1pt, black, dashed]
table {%
0 0.625
100 0.625
};
\addlegendentry{Optimal}
\end{axis}

\end{tikzpicture}
    }
    \caption{Configuration space = 704 states}
    \label{fig:exp:0:subfig:large}
\end{subfigure}%


\caption{Power (normalized to exact execution) consumption achieved by different learning algorithms provided a goal to minimize power. Ideally, the policy should learn to reduce power consumption as quickly as possible toward the minimum (black dashed line).}
\label{fig:exp:0}
\end{figure}
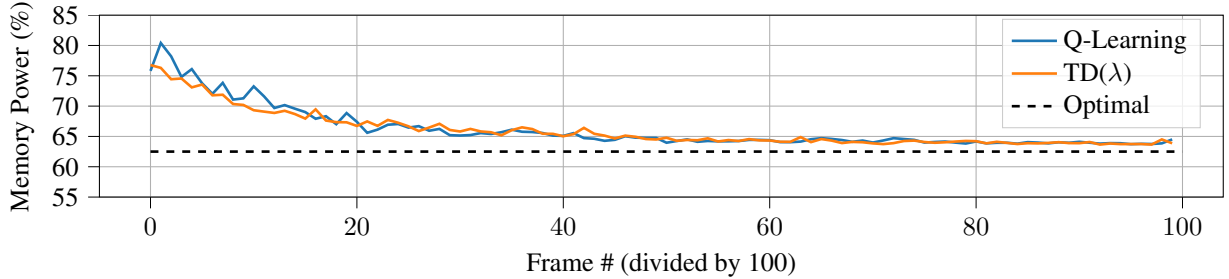
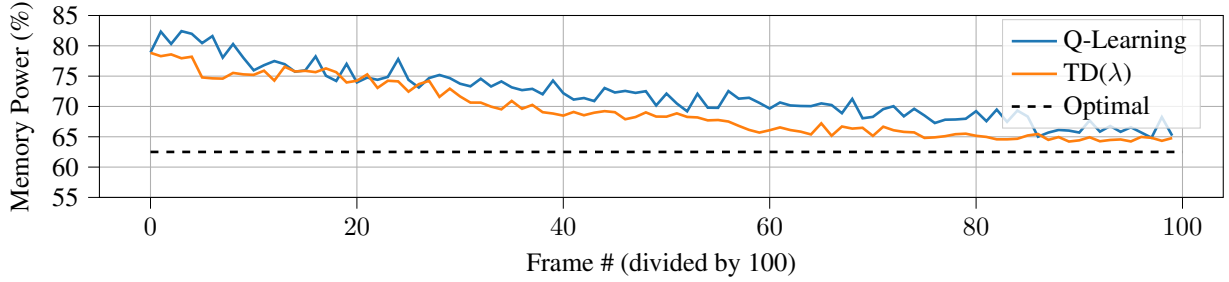

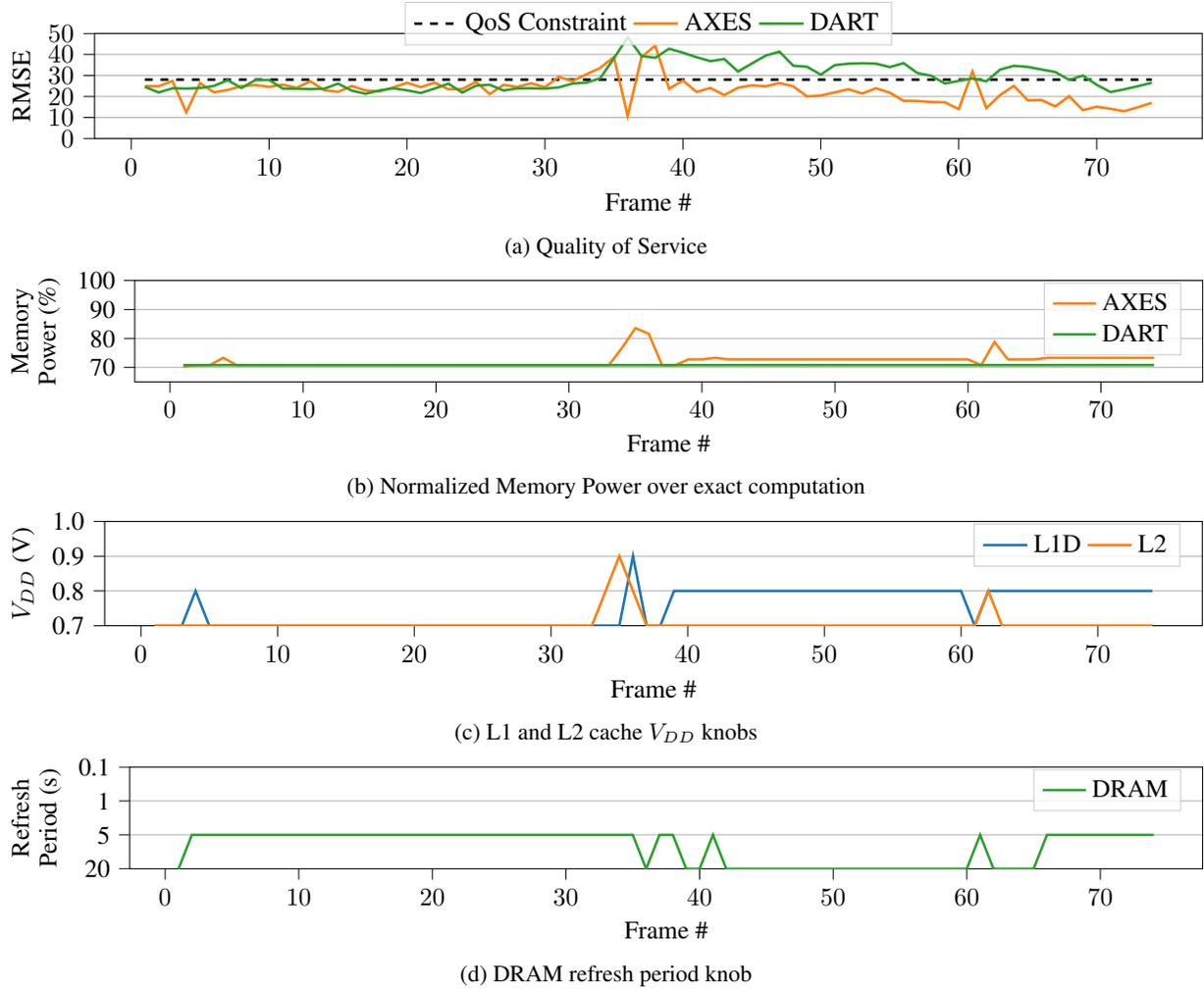
\begin{figure}[!thb]
  \begin{subfigure}[]{\linewidth}
      \centering
      \resizebox{\linewidth}{!}{
\begin{tikzpicture}

\definecolor{color0}{rgb}{0.83921568627451,0.152941176470588,0.156862745098039}
\definecolor{color1}{rgb}{1,0.498039215686275,0.0549019607843137}
\definecolor{color2}{rgb}{0.172549019607843,0.627450980392157,0.172549019607843}

\begin{axis}[
height=3cm,
legend cell align={left},
legend columns=3,
legend style={fill opacity=0.8, draw opacity=1, text opacity=1, at={(0.5,1.3)}, anchor=north, draw=white!80!black},
minor xtick={},
minor ytick={},
tick align=outside,
tick pos=left,
width=\linewidth,
x grid style={white!69.0196078431373!black},
xlabel={Frame \#},
xmin=-2.65, xmax=77.65,
xtick style={color=black},
xtick={-10,0,10,20,30,40,50,60,70,80},
y grid style={white!69.0196078431373!black},
ylabel={RMSE},
ymajorgrids,
ymin=0, ymax=50,
ytick style={color=black},
ytick={0,10,20,30,40,50}
]
\addplot [line width=1pt, black, dashed]
table {%
1 28
2 28
3 28
4 28
5 28
6 28
7 28
8 28
9 28
10 28
11 28
12 28
13 28
14 28
15 28
16 28
17 28
18 28
19 28
20 28
21 28
22 28
23 28
24 28
25 28
26 28
27 28
28 28
29 28
30 28
31 28
32 28
33 28
34 28
35 28
36 28
37 28
38 28
39 28
40 28
41 28
42 28
43 28
44 28
45 28
46 28
47 28
48 28
49 28
50 28
51 28
52 28
53 28
54 28
55 28
56 28
57 28
58 28
59 28
60 28
61 28
62 28
63 28
64 28
65 28
66 28
67 28
68 28
69 28
70 28
71 28
72 28
73 28
74 28
};
\addlegendentry{QoS Constraint}
\addplot [line width=1pt, color1]
table {%
1 24.9307011835945
2 24.8662973013127
3 27.3712118591114
4 12.4331486506564
5 26.4050732658733
6 21.9624873881063
7 23.1289347082531
8 24.904959616396
9 25.4400633201444
10 24.5809156329107
11 25.6159473338875
12 24.1199414002897
13 27.2302433071793
14 22.9898538638639
15 22.2237819548114
16 24.9564161993874
17 22.863957763928
18 22.3819603349997
19 24.2921815658305
20 26.5020619239634
21 24.4238474709471
22 26.5383412864979
23 23.4593648011833
24 23.5139850492792
25 26.9341892476112
26 21.144089764287
27 25.5532705887933
28 24.51559286095
29 26.2589186012668
30 24.3844223484389
31 29.3064283666221
32 27.3360378734614
33 30.5389997259466
34 33.5514400580898
35 38.7666924732108
36 10.7151713480021
37 38.7418658984755
38 44.1870419161031
39 23.6364191228042
40 27.5813137964571
41 22.1515093017742
42 24.0400297237611
43 20.6530134309357
44 24.159798119406
45 25.3263500268361
46 24.8146538451485
47 26.392924623286
48 24.853396497694
49 20.0062135234395
50 20.4658221361915
51 21.9039986360938
52 23.4320069320499
53 21.3853995999873
54 23.90625
55 21.8012657478924
56 17.9442149689607
57 17.8546277145132
58 17.3997738686978
59 17.2330877857302
60 13.9869339647027
61 31.8548705379356
62 14.4825981735997
63 20.6685361790353
64 25.0718073853518
65 18.1750753210581
66 18.3331988200988
67 15.2799979734351
68 20.1499725954303
69 13.4968299097839
70 15.1323616804651
71 14.1238403335299
72 12.9882452045809
73 14.8974060146694
74 16.9894320845877
};
\addlegendentry{\seams{}}
\addplot [line width=1pt, color2]
table {%
1 24.5809156329107
2 21.9039986360938
3 23.9196616420087
4 23.7717174180933
5 24.0266852710405
6 25.0077669042994
7 27.7551838549822
8 24.0266852710405
9 27.789827326576
10 27.8704946292792
11 23.7582222609353
12 23.6635406682987
13 23.5139850492792
14 23.7041648023564
15 26.1242251951221
16 22.7796404843578
17 21.2952288557096
18 22.8779804317681
19 23.8794041184268
20 22.9339853636277
21 21.6684649239186
22 23.7717174180933
23 26.111945887654
24 21.8600293740186
25 25.1484406992217
26 25.5783597166717
27 22.849926490582
28 23.852528021989
29 23.90625
30 23.7312089238757
31 24.3317561479709
32 26.1610285360209
33 26.5986969279896
34 28.6646847406074
35 38.1243596257695
36 48.193322406469
37 39.202702937599
38 38.434352929186
39 42.7107699785322
40 40.82965822496
41 38.6589955021938
42 36.8060796608386
43 37.8457369420586
44 31.8649368584706
45 35.673312535742
46 39.4229674906272
47 41.3370723580706
48 34.6425217706604
49 34.1483358647174
50 30.3810655536658
51 34.9007763064057
52 35.6103251772987
53 35.772068591783
54 35.6283329282233
55 33.9788641691754
56 35.8616108309714
57 31.173023090614
58 29.8485836161793
59 26.197780174655
60 27.4063407016225
61 28.5862586435972
62 27.2537887013814
63 32.7875987764082
64 34.5962020256888
65 34.0731191745647
66 32.8462353635362
67 31.6427380747829
68 27.9279714785928
69 29.9344172473342
70 25.5532705887933
71 22.12253413005
72 23.4046170840771
73 24.917833724062
74 26.526253678817
};
\addlegendentry{DART}
\end{axis}

\end{tikzpicture}
      }
      \caption{Quality of Service}
      \label{fig:exp:1:subfig:qos}
  \end{subfigure}%
  
  \begin{subfigure}[t]{\linewidth}
      \centering
      \resizebox{\linewidth}{!}{
\begin{tikzpicture}

\definecolor{color0}{rgb}{1,0.498039215686275,0.0549019607843137}
\definecolor{color1}{rgb}{0.172549019607843,0.627450980392157,0.172549019607843}

\begin{axis}[
height=3cm,
legend cell align={left},
legend style={fill opacity=0.8, draw opacity=1, text opacity=1, draw=white!80!black},
minor xtick={},
minor ytick={},
tick align=outside,
tick pos=left,
width=\linewidth,
x grid style={white!69.0196078431373!black},
xlabel={Frame \#},
xmin=-2.65, xmax=77.65,
xtick style={color=black},
xtick={-10,0,10,20,30,40,50,60,70,80},
y grid style={white!69.0196078431373!black},
ylabel style={align=center}, ylabel={Memory\\Power (\%)},
ymajorgrids,
ymin=0.65, ymax=1,
ytick style={color=black},
ytick={0.6,0.7,0.8,0.9,1},
yticklabels={60,70,80,90,100}
]
\addplot [line width=1pt, color0]
table {%
1 0.70252318
2 0.7080274084
3 0.7080274084
4 0.7331874084
5 0.7080274084
6 0.7080274084
7 0.7080274084
8 0.7080274084
9 0.7080274084
10 0.7080274084
11 0.7080274084
12 0.7080274084
13 0.7080274084
14 0.7080274084
15 0.7080274084
16 0.7080274084
17 0.7080274084
18 0.7080274084
19 0.7080274084
20 0.7080274084
21 0.7080274084
22 0.7080274084
23 0.7080274084
24 0.7080274084
25 0.7080274084
26 0.7080274084
27 0.7080274084
28 0.7080274084
29 0.7080274084
30 0.7080274084
31 0.7080274084
32 0.7080274084
33 0.7080274084
34 0.7683094084
35 0.8356834084
36 0.81608518
37 0.7080274084
38 0.7080274084
39 0.72768318
40 0.72768318
41 0.7331874084
42 0.72768318
43 0.72768318
44 0.72768318
45 0.72768318
46 0.72768318
47 0.72768318
48 0.72768318
49 0.72768318
50 0.72768318
51 0.72768318
52 0.72768318
53 0.72768318
54 0.72768318
55 0.72768318
56 0.72768318
57 0.72768318
58 0.72768318
59 0.72768318
60 0.72768318
61 0.7080274084
62 0.78796518
63 0.72768318
64 0.72768318
65 0.72768318
66 0.7331874084
67 0.7331874084
68 0.7331874084
69 0.7331874084
70 0.7331874084
71 0.7331874084
72 0.7331874084
73 0.7331874084
74 0.7331874084
};
\addlegendentry{\seams{}}
\addplot [line width=1pt, color1]
table {%
1 0.708030869709611
2 0.708023744064961
3 0.708026461992829
4 0.708024351220012
5 0.708029550007897
6 0.708031301792206
7 0.708015780807206
8 0.708021377979099
9 0.708029416609203
10 0.70802906543367
11 0.708018514306331
12 0.708026146584993
13 0.708023290744078
14 0.708028436552946
15 0.708026050962309
16 0.708028995135418
17 0.708021076346738
18 0.708019465538096
19 0.708022733189288
20 0.708019392431773
21 0.708028430260141
22 0.708015037787037
23 0.708018680261658
24 0.708020874712344
25 0.708011576904377
26 0.708021518356047
27 0.708017690937317
28 0.708018792597427
29 0.708014162652961
30 0.708024967938712
31 0.708014302649925
32 0.707987837982475
33 0.707987837982475
34 0.707977980218606
35 0.707963715452352
36 0.707947746088869
37 0.707932413433788
38 0.707946624411724
39 0.707955804489844
40 0.707956440718401
41 0.70794644835506
42 0.707969223787316
43 0.707980217482898
44 0.708001302521392
45 0.708052259755275
46 0.708043652376396
47 0.708054251427109
48 0.708057501990285
49 0.708028142091541
50 0.708014900038978
51 0.707992012337362
52 0.707992050999624
53 0.707992050999624
54 0.707998906762416
55 0.707998410154768
56 0.707994615292055
57 0.707989929822917
58 0.707999041192543
59 0.708012773168974
60 0.708011242178664
61 0.708023304440639
62 0.708022949031641
63 0.708011087202241
64 0.708003581063674
65 0.708009945758864
66 0.708013536479354
67 0.708042963417408
68 0.708044529646354
69 0.708044812420435
70 0.708062615445117
71 0.70809624343903
72 0.708070683816489
73 0.708070683816489
74 0.708074872987271
};
\addlegendentry{DART}
\end{axis}

\end{tikzpicture}
      }
      \caption{Normalized Memory Power over exact computation}
      \label{fig:exp:1:subfig:power}
  \end{subfigure}%
  
  \begin{subfigure}[t]{\linewidth}
  \centering
      \resizebox{\linewidth}{!}{
\begin{tikzpicture}

\definecolor{color0}{rgb}{0.12156862745098,0.466666666666667,0.705882352941177}
\definecolor{color1}{rgb}{1,0.498039215686275,0.0549019607843137}

\begin{axis}[
height=3cm,
legend cell align={left},
legend columns=2,
legend style={fill opacity=0.8, draw opacity=1, text opacity=1, draw=white!80!black},
minor xtick={},
minor ytick={},
tick align=outside,
tick pos=left,
width=\linewidth,
x grid style={white!69.0196078431373!black},
xlabel={Frame \#},
xmin=-2.65, xmax=77.65,
xtick style={color=black},
xtick={-10,0,10,20,30,40,50,60,70,80},
y grid style={white!69.0196078431373!black},
ylabel={\(\displaystyle V_{DD}\) (V)},
ymajorgrids,
ymin=1, ymax=4,
ytick style={color=black},
ytick={1,2,3,4},
yticklabels={0.7,0.8,0.9,1.0}
]
\addplot [line width=1pt, color0]
table {%
1 1
2 1
3 1
4 2
5 1
6 1
7 1
8 1
9 1
10 1
11 1
12 1
13 1
14 1
15 1
16 1
17 1
18 1
19 1
20 1
21 1
22 1
23 1
24 1
25 1
26 1
27 1
28 1
29 1
30 1
31 1
32 1
33 1
34 1
35 1
36 3
37 1
38 1
39 2
40 2
41 2
42 2
43 2
44 2
45 2
46 2
47 2
48 2
49 2
50 2
51 2
52 2
53 2
54 2
55 2
56 2
57 2
58 2
59 2
60 2
61 1
62 2
63 2
64 2
65 2
66 2
67 2
68 2
69 2
70 2
71 2
72 2
73 2
74 2
};
\addlegendentry{L1D}
\addplot [line width=1pt, color1]
table {%
1 1
2 1
3 1
4 1
5 1
6 1
7 1
8 1
9 1
10 1
11 1
12 1
13 1
14 1
15 1
16 1
17 1
18 1
19 1
20 1
21 1
22 1
23 1
24 1
25 1
26 1
27 1
28 1
29 1
30 1
31 1
32 1
33 1
34 2
35 3
36 2
37 1
38 1
39 1
40 1
41 1
42 1
43 1
44 1
45 1
46 1
47 1
48 1
49 1
50 1
51 1
52 1
53 1
54 1
55 1
56 1
57 1
58 1
59 1
60 1
61 1
62 2
63 1
64 1
65 1
66 1
67 1
68 1
69 1
70 1
71 1
72 1
73 1
74 1
};
\addlegendentry{L2}
\end{axis}

\end{tikzpicture}
      }
      \caption{L1 and L2 cache $V_{DD}$ knobs}
      \label{fig:exp:1:subfig:knobsl1l2}
  \end{subfigure}%
  
  \begin{subfigure}[t]{\linewidth}
  \centering
      \resizebox{\linewidth}{!}{
\begin{tikzpicture}

\definecolor{color0}{rgb}{0.172549019607843,0.627450980392157,0.172549019607843}

\begin{axis}[
height=3cm,
legend cell align={left},
legend style={fill opacity=0.8, draw opacity=1, text opacity=1, draw=white!80!black},
minor xtick={},
minor ytick={},
tick align=outside,
tick pos=left,
width=\linewidth,
x grid style={white!69.0196078431373!black},
xlabel={Frame \#},
xmin=-2.65, xmax=77.65,
xtick style={color=black},
xtick={-10,0,10,20,30,40,50,60,70,80},
y grid style={white!69.0196078431373!black},
ylabel style={align=center}, ylabel={Refresh\\Period (s)},
ymajorgrids,
ymin=1, ymax=4,
ytick style={color=black},
ytick={1,2,3,4},
yticklabels={20,5,1,0.1}
]
\addplot [line width=1pt, color0]
table {%
1 1
2 2
3 2
4 2
5 2
6 2
7 2
8 2
9 2
10 2
11 2
12 2
13 2
14 2
15 2
16 2
17 2
18 2
19 2
20 2
21 2
22 2
23 2
24 2
25 2
26 2
27 2
28 2
29 2
30 2
31 2
32 2
33 2
34 2
35 2
36 1
37 2
38 2
39 1
40 1
41 2
42 1
43 1
44 1
45 1
46 1
47 1
48 1
49 1
50 1
51 1
52 1
53 1
54 1
55 1
56 1
57 1
58 1
59 1
60 1
61 2
62 1
63 1
64 1
65 1
66 2
67 2
68 2
69 2
70 2
71 2
72 2
73 2
74 2
};
\addlegendentry{DRAM}
\end{axis}

\end{tikzpicture}
      }
      \caption{DRAM refresh period knob}
      \label{fig:exp:1:subfig:knobsdram}
  \end{subfigure}%
  \caption{\seams{} self-optimizing power within quality constraint}
  \label{fig:exp:1}
  \end{figure}
  
Without any QoS constraints, approximation knobs should be set to the configuration corresponding to the lowest power. 
The goal of a policy should be to reach the optimal configuration as quickly as possible. 
Figure \ref{fig:exp:0} shows the comparison of the two methods, along with the optimal configuration corresponding to maximum energy savings. 
The plots are averaged over 16 runs to remove the effect of any outliers. 

\begin{figure}[thb]
\begin{subfigure}[]{\linewidth}
    \centering
    \resizebox{\linewidth}{!}{
\begin{tikzpicture}

\definecolor{color0}{rgb}{0.83921568627451,0.152941176470588,0.156862745098039}
\definecolor{color1}{rgb}{1,0.498039215686275,0.0549019607843137}

\begin{axis}[
height=3cm,
legend cell align={left},
legend columns=3,
legend style={fill opacity=0.8, draw opacity=1, text opacity=1, at={(0.5,1.3)}, anchor=north, draw=white!80!black},
minor xtick={},
minor ytick={},
tick align=outside,
tick pos=left,
width=\linewidth,
x grid style={white!69.0196078431373!black},
xlabel={Frame \#},
xmin=-3.9, xmax=103.9,
xtick style={color=black},
xtick={0,10,20,30,40,50,60,70,80,90,100},
y grid style={white!69.0196078431373!black},
ylabel={RMSE},
ymajorgrids,
ymin=0, ymax=100,
ytick style={color=black},
ytick={0,20,40,60,80,100}
]
\addplot [line width=1pt, black, dashed]
table {%
1 10
2 10
3 10
4 10
5 10
6 10
7 10
8 10
9 10
10 10
11 10
12 10
13 10
14 10
15 10
16 10
17 10
18 10
19 10
20 10
21 10
22 10
23 10
24 10
25 10
26 10
27 10
28 10
29 10
30 85
31 85
32 85
33 85
34 85
35 85
36 85
37 85
38 85
39 85
40 85
41 85
42 85
43 85
44 85
45 85
46 85
47 85
48 85
49 85
50 85
51 85
52 85
53 85
54 85
55 85
56 85
57 85
58 85
59 85
60 30
61 30
62 30
63 30
64 30
65 30
66 30
67 30
68 30
69 30
70 30
71 30
72 30
73 30
74 30
75 30
76 30
77 30
78 30
79 30
80 30
81 30
82 30
83 30
84 30
85 30
86 30
87 30
88 30
89 30
90 30
91 30
92 30
93 30
94 30
95 30
96 30
97 30
98 30
99 30
};
\addlegendentry{QoS Constraint}
\addplot [line width=1pt, color1]
table {%
1 6.70072237879822
2 4.53051522255672
3 0
4 6.40711607227748
5 3.58168676933597
6 8.08858755157818
7 7.88784688943291
8 7.42714376383253
9 6.55556316274062
10 6.50645712926782
11 6.04658319415715
12 6.55556316274062
13 7.96875
14 6.60430408039815
15 8.84611376216595
16 2.26525761127836
17 4.66994820026752
18 4.0837512347996
19 2.11895446854639
20 4.31292199845682
21 6.40711607227748
22 8.1675024695992
23 10.534056815062
24 13.8718132089697
25 4.73812623289667
26 4.23790893709279
27 0
28 7.63999898671757
29 6.84280296477786
30 9.33989640053503
31 21.7275877211857
32 76.5677178184894
33 77.5002863814357
34 77.9253557052502
35 75.2921334683505
36 76.1898106797783
37 75.7677032174809
38 76.4838998209308
39 76.9062465368844
40 76.1982289783156
41 76.0929333071723
42 76.3622003065221
43 78.6545264736947
44 78.5198544485001
45 78.6749112158987
46 78.744179866998
47 78.638214875329
48 78.2088186727163
49 77.913007834204
50 79.1463634002766
51 79.9927864290388
52 77.7440570432221
53 79.4860142161243
54 79.7558892141928
55 80.3248658458476
56 80.177000348873
57 80.7787497415077
58 79.8242198361073
59 81.8436979329211
60 80.5322179096594
61 18.9863185265912
62 20.7768706937273
63 24.3975711346996
64 26.9103642192887
65 24.5025074087976
66 27.5696834717585
67 24.2393149491595
68 27.1240361030037
69 28.099696635272
70 28.4851078142158
71 21.3253281341322
72 24.7758506097672
73 25.0590123820254
74 3.67013679844746
75 12.814232144555
76 3.84093115411899
77 6.84280296477786
78 7.25236295740286
79 8.51357226092816
80 23.5139850492792
81 24.0400297237611
82 24.080018711296
83 22.9758994748005
84 22.4534912311184
85 25.7966318960065
86 27.8704946292792
87 25.2248412014102
88 24.3053802527949
89 23.9196616420087
90 23.7582222609353
91 26.6829654706297
92 23.3222545397777
93 25.4274536225567
94 24.5548073773937
95 25.1866699192116
96 25.1739332964405
97 24.2128383547485
98 26.1364987335935
99 25.5783597166717
};
\addlegendentry{\seams{}}
\end{axis}

\end{tikzpicture}
    }
    \caption{Quality of Service}
    \label{fig:exp:2:subfig:qos}
\end{subfigure}%

\begin{subfigure}[t]{\linewidth}
    \centering
    \resizebox{\linewidth}{!}{
\begin{tikzpicture}

\definecolor{color0}{rgb}{1,0.498039215686275,0.0549019607843137}

\begin{axis}[
height=3cm,
legend cell align={left},
legend style={fill opacity=0.8, draw opacity=1, text opacity=1, draw=white!80!black},
minor xtick={},
minor ytick={},
tick align=outside,
tick pos=left,
width=\linewidth,
x grid style={white!69.0196078431373!black},
xlabel={Frame \#},
xmin=-3.9, xmax=103.9,
xtick style={color=black},
xtick={0,10,20,30,40,50,60,70,80,90,100},
y grid style={white!69.0196078431373!black},
ylabel style={align=center}, ylabel={Memory\\Power (\%)},
ymajorgrids,
ymin=0.65, ymax=1,
ytick style={color=black},
ytick={0.6,0.7,0.8,0.9,1},
yticklabels={60,70,80,90,100}
]
\addplot [line width=1pt, color0]
table {%
1 0.862624855421537
2 0.821512637496778
3 0.888962661681621
4 0.787968609720017
5 0.787965341496551
6 0.787966615279827
7 0.787966444979445
8 0.787967428066824
9 0.787964101639083
10 0.787969155933958
11 0.79346850948787
12 0.793469995566885
13 0.793471037603216
14 0.793470736892715
15 0.793472757813075
16 0.793472013698347
17 0.793472780926813
18 0.793469547664132
19 0.793470614088914
20 0.793472378729533
21 0.793470692895775
22 0.793470496842036
23 0.761175510212322
24 0.761172675883362
25 0.86262091783994
26 0.821529451947091
27 0.888970415401984
28 0.787967744584853
29 0.787966782529842
30 0.793470407132343
31 0.702527736600235
32 0.674368959522315
33 0.674367081073089
34 0.674388357332947
35 0.674375977259176
36 0.67437946390358
37 0.674266469063131
38 0.674375604975087
39 0.674378277942623
40 0.674379506290314
41 0.674385394056208
42 0.674380700301706
43 0.67437387162652
44 0.674392817259832
45 0.67438833623838
46 0.67438386639868
47 0.674379842008151
48 0.674388782727615
49 0.674375851657619
50 0.674388604371971
51 0.674379293230064
52 0.6743960702769
53 0.67438190301843
54 0.67439037015995
55 0.674389002236779
56 0.674378255344529
57 0.674382871485042
58 0.674379790658958
59 0.674274461295398
60 0.674406180215627
61 0.708024686541372
62 0.768365800879982
63 0.702517661892297
64 0.785130193399643
65 0.702525914073247
66 0.785128943341373
67 0.880628935763556
68 0.852572352885317
69 0.880623753464302
70 0.880625466715364
71 0.880621257570021
72 0.813196385606117
73 0.852560785328004
74 0.787965954909581
75 0.787965990113838
76 0.787965053984925
77 0.787967547726141
78 0.787964448019474
79 0.787964882420728
80 0.70253830102682
81 0.70252955468554
82 0.702516801006656
83 0.702533534929898
84 0.702522369253701
85 0.777545095564111
86 0.785129680348724
87 0.785130807279045
88 0.785131888696734
89 0.785129492852387
90 0.78513081534351
91 0.785130806642747
92 0.813192103836627
93 0.813196155051454
94 0.813199574853041
95 0.813188229455527
96 0.813191026009359
97 0.813201923208813
98 0.852558697096218
99 0.852574197734837
};
\addlegendentry{\seams{}}
\end{axis}

\end{tikzpicture}
    }
    \caption{Normalized Memory Power over exact computation}
    \label{fig:exp:2:subfig:power}
\end{subfigure}%

\begin{subfigure}[t]{\linewidth}
\centering
    \resizebox{\linewidth}{!}{
\begin{tikzpicture}

\definecolor{color0}{rgb}{0.12156862745098,0.466666666666667,0.705882352941177}
\definecolor{color1}{rgb}{1,0.498039215686275,0.0549019607843137}

\begin{axis}[
height=3cm,
legend cell align={left},
legend columns=2,
legend style={fill opacity=0.8, draw opacity=1, text opacity=1,  at={(0.5,1.1)}, anchor=north, draw=white!80!black},
minor xtick={},
minor ytick={},
tick align=outside,
tick pos=left,
width=\linewidth,
x grid style={white!69.0196078431373!black},
xlabel={Frame \#},
xmin=-3.9, xmax=103.9,
xtick style={color=black},
xtick={0,10,20,30,40,50,60,70,80,90,100},
y grid style={white!69.0196078431373!black},
ylabel={\(\displaystyle V_{DD}\) (V)},
ymajorgrids,
ymin=1, ymax=4,
ytick style={color=black},
ytick={1,2,3,4},
yticklabels={0.7,0.8,0.9,1.0}
]
\addplot [line width=1pt, color0]
table {%
1 2
2 3
3 3
4 2
5 2
6 2
7 2
8 2
9 2
10 2
11 2
12 2
13 2
14 2
15 2
16 2
17 2
18 2
19 2
20 2
21 2
22 2
23 3
24 3
25 2
26 3
27 3
28 2
29 2
30 2
31 1
32 2
33 2
34 2
35 2
36 2
37 2
38 2
39 2
40 2
41 2
42 2
43 2
44 2
45 2
46 2
47 2
48 2
49 2
50 2
51 2
52 2
53 2
54 2
55 2
56 2
57 2
58 2
59 2
60 2
61 1
62 1
63 1
64 2
65 1
66 2
67 3
68 2
69 3
70 3
71 3
72 3
73 2
74 2
75 2
76 2
77 2
78 2
79 2
80 1
81 1
82 1
83 1
84 1
85 1
86 2
87 2
88 2
89 2
90 2
91 2
92 3
93 3
94 3
95 3
96 3
97 3
98 2
99 2
};
\addlegendentry{L1D}
\addplot [line width=1pt, color1]
table {%
1 2
2 2
3 3
4 2
5 2
6 2
7 2
8 2
9 2
10 2
11 2
12 2
13 2
14 2
15 2
16 2
17 2
18 2
19 2
20 2
21 2
22 2
23 1
24 1
25 2
26 2
27 3
28 2
29 2
30 2
31 1
32 0
33 0
34 0
35 0
36 0
37 0
38 0
39 0
40 0
41 0
42 0
43 0
44 0
45 0
46 0
47 0
48 0
49 0
50 0
51 0
52 0
53 0
54 0
55 0
56 0
57 0
58 0
59 0
60 0
61 1
62 2
63 1
64 2
65 1
66 2
67 3
68 3
69 3
70 3
71 3
72 2
73 3
74 2
75 2
76 2
77 2
78 2
79 2
80 1
81 1
82 1
83 1
84 1
85 1
86 2
87 2
88 2
89 2
90 2
91 2
92 2
93 2
94 2
95 2
96 2
97 2
98 3
99 3
};
\addlegendentry{L2}
\end{axis}

\end{tikzpicture}
    }
    \caption{L1 and L2 cache $V_{DD}$ knobs}
    \label{fig:exp:2:subfig:knobsl1l2}
\end{subfigure}%

\begin{subfigure}[t]{\linewidth}
\centering
    \resizebox{\linewidth}{!}{
\begin{tikzpicture}

\definecolor{color0}{rgb}{0.172549019607843,0.627450980392157,0.172549019607843}

\begin{axis}[
height=3cm,
legend cell align={left},
legend style={fill opacity=0.8, draw opacity=1, text opacity=1, draw=white!80!black},
minor xtick={},
minor ytick={},
tick align=outside,
tick pos=left,
width=\linewidth,
x grid style={white!69.0196078431373!black},
xlabel={Frame \#},
xmin=-3.9, xmax=103.9,
xtick style={color=black},
xtick={0,10,20,30,40,50,60,70,80,90,100},
y grid style={white!69.0196078431373!black},
ylabel style={align=center}, ylabel={Refresh\\Period (s)},
ymajorgrids,
ymin=1, ymax=4,
ytick style={color=black},
ytick={1,2,3,4},
yticklabels={20,5,1,0.1}
]
\addplot [line width=1pt, color0]
table {%
1 3
2 2
3 2
4 1
5 1
6 1
7 1
8 1
9 1
10 1
11 2
12 2
13 2
14 2
15 2
16 2
17 2
18 2
19 2
20 2
21 2
22 2
23 2
24 2
25 3
26 2
27 2
28 1
29 1
30 2
31 1
32 1
33 1
34 1
35 1
36 1
37 1
38 1
39 1
40 1
41 1
42 1
43 1
44 1
45 1
46 1
47 1
48 1
49 1
50 1
51 1
52 1
53 1
54 1
55 1
56 1
57 1
58 1
59 1
60 1
61 2
62 2
63 1
64 0
65 1
66 0
67 0
68 0
69 0
70 0
71 0
72 0
73 0
74 1
75 1
76 1
77 1
78 1
79 1
80 1
81 1
82 1
83 1
84 1
85 1
86 0
87 0
88 0
89 0
90 0
91 0
92 0
93 0
94 0
95 0
96 0
97 0
98 0
99 0
85 0
};
\addlegendentry{DRAM}
\end{axis}

\end{tikzpicture}
    }
    \caption{DRAM refresh period knob}
    \label{fig:exp:2:subfig:knobsdram}
\end{subfigure}%
\caption{\seams{} self-adapting to user-specified quality constraints by coordination across the memory hierarchy.}
\label{fig:exp:2}
\end{figure}
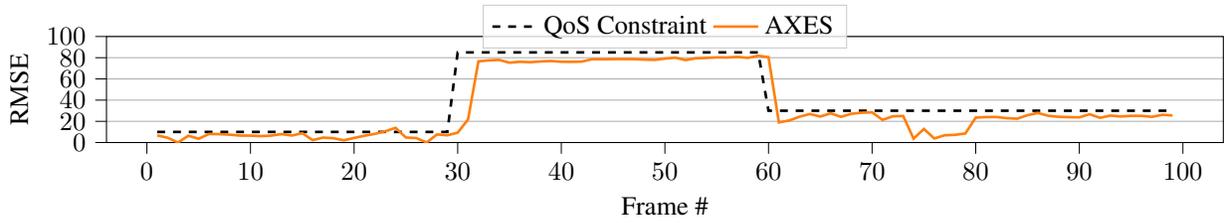
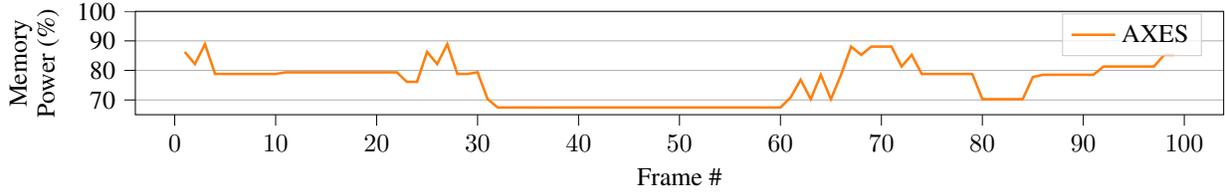
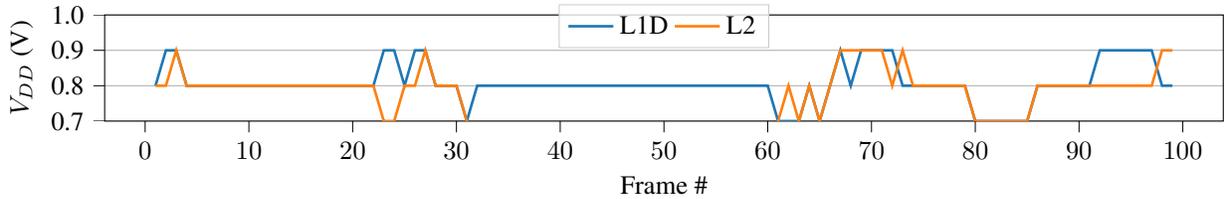
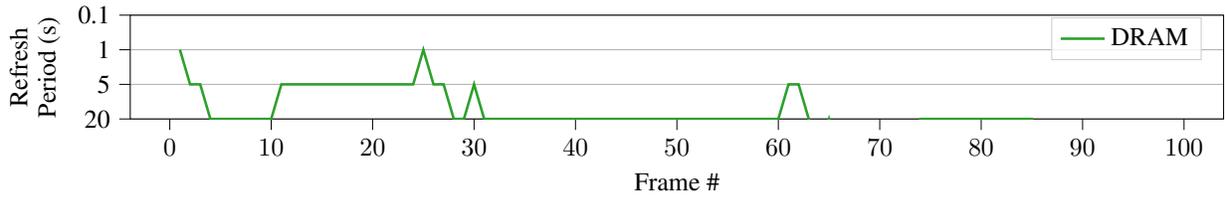
The x-axis in Figure \ref{fig:exp:0} represents frames processed, and the y-axis represents average memory power (normalized to the exact execution) for each episode for canny edge detection application\cite{Canny}. 
We make two major observations.
First, both algorithms can eventually converge to the optimal policy.
Second, in Figure \ref{fig:exp:0:subfig:small}, when the configuration space is small (i.e., we restrict the allowable knob settings), both Q-learning and TD($\lambda$) converge at an equal rate. 
However, when configuration space is increased (Figure \ref{fig:exp:0:subfig:large}), TD($\lambda$) can improve its policy faster than Q-learning because it uses short-term memory in the form of eligibility traces. The traces are used to update multiple state-action pairs based on the reward obtained, instead of just one state-action pair in Q-learning at every step.
We conclude that with growing complexity in configuration knobs, TD($\lambda$) is the better algorithm.
Thus, for the rest of the experiments, we only use the TD($\lambda$) algorithm in \seams{}.

\subsection{Self-optimization}

To show that \seams{} is capable of self-optimizing the approximation knobs in the memory hierarchy within the QoS budget specified by the application, we study \seams{}' behavior for unexperienced inputs.
We expose an agent with an established policy to varying inputs and compare it to state-of-art approximation management policy DART \cite{Yarmand2020}. 
DART is a design-time technique that uses a branch and bound algorithm to consider the worst-case effects of all possible approximation knob configurations for a memory hierarchy. 
We train DART on a set of scenes used during the policy initialization phase. 
The goal is to honor the QoS constraint specified by the application while maximizing energy-efficiency. 
In our case, this means keeping the RMSE below a specified value.
For a QoS constraint of 10 RMSE DART statically sets the knobs as follows: L1 $V_{DD}$:\SI{0.8}{\volt}, L2 $V_{DD}$:\SI{0.8}{\volt} and DRAM $T_{REF}$:\SI{0.5}{\second}.
The energy/frame reported in all results is normalized with respect to the the knob configuration of L1 $V_{DD}$:\SI{1}{\volt}, L2 $V_{DD}$:\SI{1}{\volt} and DRAM $T_{REF}$:\SI{0.1}{\second}. 

In Figure \ref{fig:exp:1:subfig:qos}, a QoS constraint of 28 (RMSE) has been specified by the user, which is marked with a black line. 
The key observations here are as follows:
(1) Frame 32 is a key-frame where a scene change occurs. 
Neither of the tested policies has experienced this scene previously, and the scene requires a new configuration of knobs to remain to meet the QoS requirement. 
In frame 32, DART immediately violates the QoS constraint and continues to do so.
\seams{} can take actions and reach a new configuration while remaining within the quality constraint.
Initially, when \seams{} detects there is an overshoot, it penalizes the current action and takes action to reduce the QoS error. 
This leads to a conservative state, with more room for QoS relaxation. 
Then, \seams{} increases the degree of approximation. In the subsequent cycles \seams{} self-optimizes until it reaches a stable state.
(2) Figure \ref{fig:exp:1:subfig:power} shows the average power for each frame. 
There is no significant change in power with DART because it uses a fixed knob configuration at runtime.
The normalized energy/frame required by \seams{} is \SI{72.3}{\%}, and normalized energy/frame required by DART is \SI{70.8}{\%}. 
(3) We define QoS overshoot as the area under curve for the regions of QoS violations during execution. For DART, QoS overshoot is 200 versus 50 for \seams{}. Thus, DART violates the QoS requirement $4\times$ more than \seams{} (Figure \ref{fig:exp:1:subfig:qos}).
This means that \seams{} can reduce QoS violations by \SI{75}{\%} with $<$\SI{5}{\%} additional energy.
These experiments demonstrate that \seams{} is self-optimizing, i.e., can continuously learn configurations that meet the QoS constraint when exposed to unknown inputs.

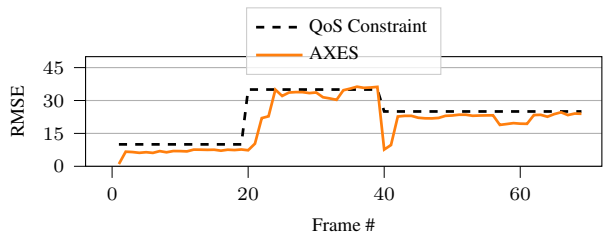
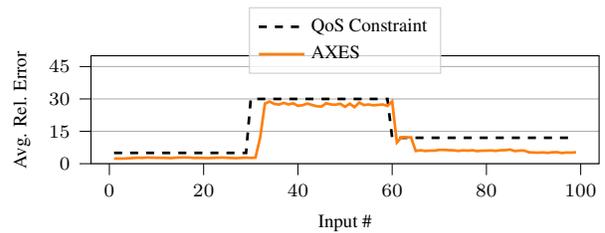
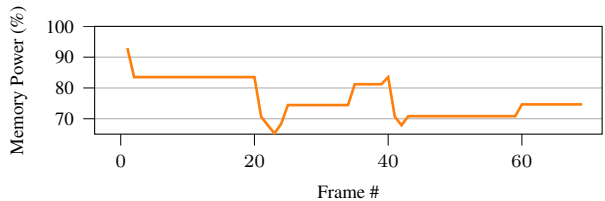
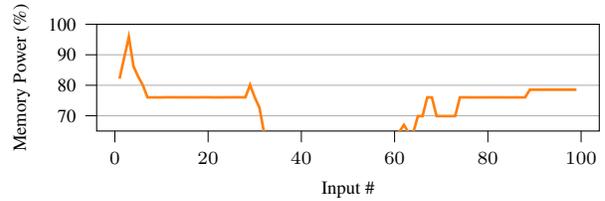
\begin{figure}
  \centering
  \begin{subfigure}[]{0.5\linewidth}
      \centering
      \resizebox{\linewidth}{!}{
\begin{tikzpicture}

\definecolor{color0}{rgb}{1,0.498039215686275,0.0549019607843137}

\begin{axis}[
font=\scriptsize,
height=3cm,
legend cell align={left},
legend columns=1,
legend style={fill opacity=0.8, draw opacity=1, text opacity=1, at={(0.5,1.45)}, anchor=north, draw=white!80!black},
minor xtick={},
minor ytick={},
tick align=outside,
tick pos=left,
width=\linewidth,
x grid style={white!69.0196078431373!black},
xmin=-3.9, xmax=72,
xtick style={color=black},
xtick={-20,0,20,40,60,80,100,120},
xlabel={Frame \#},
y grid style={white!69.0196078431373!black},
ylabel={RMSE},
ymajorgrids,
ymin=0, ymax=50,
ytick style={color=black},
ytick={0,15,30,45},
ylabel near ticks,
    xlabel near ticks,
]
\addplot [line width=1pt, black, dashed]
table {%
1 10
2 10
3 10
4 10
5 10
6 10
7 10
8 10
9 10
10 10
11 10
12 10
13 10
14 10
15 10
16 10
17 10
18 10
19 10
20 35
21 35
22 35
23 35
24 35
25 35
26 35
27 35
28 35
29 35
30 35
31 35
32 35
33 35
34 35
35 35
36 35
37 35
38 35
39 35
40 25
41 25
42 25
43 25
44 25
45 25
46 25
47 25
48 25
49 25
50 25
51 25
52 25
53 25
54 25
55 25
56 25
57 25
58 25
59 25
60 25
61 25
62 25
63 25
64 25
65 25
66 25
67 25
68 25
69 25
};
\addlegendentry{QoS Constraint}
\addplot [line width=1pt, color0]
table {%
1 1.03165
2 6.6654
3 6.4951
4 6.14525
5 6.4212
6 6.11505
7 6.9209
8 6.3858
9 6.95725
10 6.924
11 6.797
12 7.60605
13 7.57565
14 7.49745
15 7.55955
16 7.11765
17 7.5634
18 7.408
19 7.6659
20 7.3118
21 10.32045
22 21.9793
23 22.79085
24 34.9478
25 32.07045
26 33.63265
27 33.8301
28 33.8034
29 33.3731
30 33.66325
31 31.47985
32 30.946
33 30.30845
34 34.6879
35 35.42135
36 36.30315
37 35.775
38 35.9286
39 36.215
40 7.64215
41 9.69485
42 22.70815
43 22.9972
44 23.0062
45 22.14185
46 21.87715
47 21.84145
48 22.0504
49 23.01035
50 23.1614
51 23.52535
52 23.51235
53 23.0361
54 23.15125
55 23.21125
56 23.2477
57 18.87625
58 19.2677
59 19.6833
60 19.4361
61 19.37375
62 23.3632
63 23.4732
64 22.60035
65 23.7978
66 24.6244
67 23.35485
68 24.0449
69 23.94975
};
\addlegendentry{\seams{}}
\end{axis}

\end{tikzpicture}
      }
      \caption{\texttt{kmeans}:QoS}
      \label{fig:exp:subfig:kmns:qos}
  \end{subfigure}%
  ~ 
  \begin{subfigure}[]{0.5\linewidth}
      \centering
      \resizebox{\linewidth}{!}{
\begin{tikzpicture}

\definecolor{color0}{rgb}{1,0.498039215686275,0.0549019607843137}

\begin{axis}[
font=\scriptsize,
height=3cm,
legend cell align={left},
legend columns=1,
legend style={fill opacity=0.8, draw opacity=1, text opacity=1, at={(0.5,1.45)}, anchor=north, draw=white!80!black},
minor xtick={},
minor ytick={},
tick align=outside,
tick pos=left,
width=\linewidth,
x grid style={white!69.0196078431373!black},
xmin=-3.9, xmax=103.9,
xtick style={color=black},
xtick={-20,0,20,40,60,80,100,120},
xlabel={Input \#},
y grid style={white!69.0196078431373!black},
ylabel={Avg. Rel. Error},
ymajorgrids,
ymin=0, ymax=50,
ytick style={color=black},
ytick={0,15,30,45},
ylabel near ticks,
    xlabel near ticks,
]
\addplot [line width=1pt, black, dashed]
table {%
1 5
2 5
3 5
4 5
5 5
6 5
7 5
8 5
9 5
10 5
11 5
12 5
13 5
14 5
15 5
16 5
17 5
18 5
19 5
20 5
21 5
22 5
23 5
24 5
25 5
26 5
27 5
28 5
29 5
30 30
31 30
32 30
33 30
34 30
35 30
36 30
37 30
38 30
39 30
40 30
41 30
42 30
43 30
44 30
45 30
46 30
47 30
48 30
49 30
50 30
51 30
52 30
53 30
54 30
55 30
56 30
57 30
58 30
59 30
60 12
61 12
62 12
63 12
64 12
65 12
66 12
67 12
68 12
69 12
70 12
71 12
72 12
73 12
74 12
75 12
76 12
77 12
78 12
79 12
80 12
81 12
82 12
83 12
84 12
85 12
86 12
87 12
88 12
89 12
90 12
91 12
92 12
93 12
94 12
95 12
96 12
97 12
98 12
99 12
};
\addlegendentry{QoS Constraint}
\addplot [line width=1pt, color0]
table {%
1 2.392728
2 2.377065
3 2.377069
4 2.499344
5 2.605047
6 2.765629
7 2.697067
8 2.850173
9 2.785901
10 2.724666
11 2.726094
12 2.661312
13 2.571434
14 2.773077
15 2.841467
16 2.884256
17 2.834158
18 2.651703
19 2.679667
20 2.634872
21 2.586695
22 2.690217
23 2.774887
24 2.842239
25 2.723593
26 2.616246
27 2.574581
28 2.695288
29 2.795986
30 2.668373
31 2.785279
32 12.172176
33 27.82428
34 28.812234
35 27.688057
36 27.380358
37 28.186071
38 27.432465
39 27.992616
40 26.796509
41 27.101046
42 27.8663
43 27.209071
44 26.598229
45 26.468691
46 28.024152
47 27.419029
48 27.313585
49 27.788933
50 26.338114
51 27.824834
52 26.1851
53 28.335787
54 27.171896
55 27.426607
56 27.044132
57 27.22952
58 27.441193
59 26.749322
60 28.897368
61 9.794399
62 12.299802
63 12.238578
64 12.20489
65 5.897524
66 6.220998
67 5.900346
68 6.045118
69 6.053022
70 6.395673
71 6.38197
72 6.271232
73 6.198039
74 5.91295
75 6.185125
76 6.040284
77 6.234193
78 5.842802
79 6.014001
80 6.039195
81 6.139668
82 5.92878
83 6.224558
84 6.256509
85 6.543393
86 5.851694
87 6.07397
88 6.079246
89 5.272721
90 5.16946
91 5.09463
92 5.211537
93 5.017524
94 5.276655
95 5.334307
96 4.93397
97 5.179316
98 5.121304
99 5.310211
};
\addlegendentry{\seams{}}
\end{axis}

\end{tikzpicture}
      }
      \caption{\texttt{blackscholes}: QoS}
      \label{fig:exp:subfig:blk:qos}
  \end{subfigure}%
  
  \begin{subfigure}[t]{0.5\linewidth}
      \centering
      \resizebox{\linewidth}{!}{
\begin{tikzpicture}

\definecolor{color0}{rgb}{1,0.498039215686275,0.0549019607843137}

\begin{axis}[
font=\scriptsize,
height=3cm,
legend cell align={left},
legend style={fill opacity=0.8, draw opacity=1, text opacity=1, draw=white!80!black},
minor xtick={},
minor ytick={},
tick align=outside,
tick pos=left,
width=\linewidth,
x grid style={white!69.0196078431373!black},
xmin=-3.9, xmax=72,
xtick style={color=black},
xtick={-20,0,20,40,60,80,100,120},
xlabel={Frame \#},
y grid style={white!69.0196078431373!black},
ylabel={Memory Power (\%)},
ymajorgrids,
ymin=0.65, ymax=1,
ytick style={color=black},
ytick={0.6,0.7,0.8,0.9,1},
yticklabels={60,70,80,90,100},
ylabel near ticks,
    xlabel near ticks,
]
\addplot [line width=1pt, color0]
table {%
1 0.929690558985804
2 0.835250833940944
3 0.835250833940944
4 0.835250833940944
5 0.835250833940944
6 0.835250833940944
7 0.835250833940944
8 0.835250833940944
9 0.835250833940944
10 0.835250833940944
11 0.835250833940944
12 0.835250833940944
13 0.835250833940944
14 0.835250833940944
15 0.835250833940944
16 0.835250833940944
17 0.835250833940944
18 0.835250833940944
19 0.835250833940944
20 0.835250833940944
21 0.706564035865398
22 0.678964935074507
23 0.652723377408449
24 0.6834097202777
25 0.744429133195575
26 0.744429133195575
27 0.744429133195575
28 0.744429133195575
29 0.744429133195575
30 0.744429133195575
31 0.744429133195575
32 0.744429133195575
33 0.744429133195575
34 0.744429133195575
35 0.812096518353245
36 0.812096518353245
37 0.812096518353245
38 0.812096518353245
39 0.812096518353245
40 0.835250833940944
41 0.706564035865398
42 0.678964935074507
43 0.708293734818926
44 0.708293734818926
45 0.708293734818926
46 0.708293734818926
47 0.708293734818926
48 0.708293734818926
49 0.708293734818926
50 0.708293734818926
51 0.708293734818926
52 0.708293734818926
53 0.708293734818926
54 0.708293734818926
55 0.708293734818926
56 0.708293734818926
57 0.708293734818926
58 0.708293734818926
59 0.708293734818926
60 0.746632320232177
61 0.746632320232177
62 0.746632320232177
63 0.746632320232177
64 0.746632320232177
65 0.746632320232177
66 0.746632320232177
67 0.746632320232177
68 0.746632320232177
69 0.746632320232177
};
\end{axis}

\end{tikzpicture}
      }
      \caption{\texttt{kmeans}: Memory Power}
      \label{fig:exp:subfig:kmns:power}
  \end{subfigure}%
  ~ 
  \begin{subfigure}[t]{0.5\linewidth}
      \centering
      \resizebox{\linewidth}{!}{
\begin{tikzpicture}

\definecolor{color0}{rgb}{1,0.498039215686275,0.0549019607843137}

\begin{axis}[
font=\scriptsize,
height=3cm,
legend style={fill opacity=0.8, draw opacity=1, text opacity=1, draw=white!80!black},
minor xtick={},
minor ytick={},
tick align=outside,
tick pos=left,
width=\linewidth,
x grid style={white!69.0196078431373!black},
xmin=-3.9, xmax=103.9,
xtick style={color=black},
xtick={-20,0,20,40,60,80,100,120},
xlabel={Input \#},
y grid style={white!69.0196078431373!black},
ylabel={Memory Power (\%)},
ymajorgrids,
ymin=0.65, ymax=1,
ytick style={color=black},
ytick={0.6,0.7,0.8,0.9,1},
yticklabels={60,70,80,90,100},
ylabel near ticks,
xlabel near ticks,
]
\addplot [line width=1pt, color0, forget plot]
table {%
1 0.821461099631789
2 0.890384282527351
3 0.958672793729092
4 0.86191495727394
5 0.828449173284382
6 0.800539567495663
7 0.760259594661692
8 0.760384920294157
9 0.760154020243157
10 0.760262730644837
11 0.760568622730399
12 0.760588025107352
13 0.760563036971985
14 0.760422992917947
15 0.760358910124605
16 0.760661043771818
17 0.760266541727265
18 0.76052409316153
19 0.760699326082312
20 0.760637562758513
21 0.760297954031992
22 0.760404734786813
23 0.760292011831895
24 0.76059026181378
25 0.760510231598035
26 0.760633337807534
27 0.760601616346937
28 0.760483499597496
29 0.801175437295033
30 0.760427294048058
31 0.726927951317188
32 0.64476281590522
33 0.62544334540123
34 0.630793490715995
35 0.630783610297277
36 0.630883669376935
37 0.630827254116471
38 0.630869243007683
39 0.630751451645189
40 0.630893223150328
41 0.630791985593689
42 0.630919124060455
43 0.630710078028877
44 0.630827805239034
45 0.630680736243531
46 0.630787683059352
47 0.630878865135714
48 0.63085876638874
49 0.630800494772585
50 0.63079802589453
51 0.630909412498253
52 0.63076353910609
53 0.630753988307789
54 0.630734052718752
55 0.630810171281225
56 0.631516228716643
57 0.630870616500626
58 0.630769123921966
59 0.631092674011907
60 0.630735391174765
61 0.647568627754943
62 0.670086263426053
63 0.644789103048443
64 0.644870909705185
65 0.698893365830975
66 0.698905851761429
67 0.760427187320055
68 0.760225152820348
69 0.698918601095275
70 0.698921261643771
71 0.698905708599949
72 0.69888224964797
73 0.699023176256708
74 0.760345644147689
75 0.760472705102201
76 0.760266899269833
77 0.760171111383495
78 0.760284546071071
79 0.760337040475937
80 0.760218011389243
81 0.760356413044742
82 0.760434138059052
83 0.760271544218498
84 0.760259759037468
85 0.760133586926687
86 0.760344518462025
87 0.760249042019752
88 0.760282097773544
89 0.785452847376973
90 0.785424963924801
91 0.785439593440256
92 0.785403697340031
93 0.785454483612064
94 0.785434461652431
95 0.785464160238424
96 0.785448325914504
97 0.785451716537854
98 0.785461342448759
99 0.785457513983249
};
\end{axis}

\end{tikzpicture}
      }
      \caption{\texttt{blackscholes}: Memory Power}
      \label{fig:exp:subfig:blk:power}
  \end{subfigure}%
  \caption{Additional workloads.}
  \label{fig:exp:additional}
\end{figure}

\subsection{Coordination}
To show that \seams{} is capable of coordinating interdependent memory knobs, we study \seams{}' behavior when exposed to varying quality constraints. 
The goal is to adapt to new targets specified by the user and reconfigure the knobs to save more energy while processing the frames. 
Figure \ref{fig:exp:2} shows how \seams{} behaves in the scenario described.
The key observations here are as follows: 
(1) Figure \ref{fig:exp:2:subfig:qos} shows measured QoS compared to the dynamic QoS constraint. 
Initially, the QoS constraint is 10. 
At frame 30, it is relaxed and updated to 85, thus exposing an opportunity to conserve energy. 
Again, at frame 60, the constraint is changed to 30.
\seams{} can self-adapt to find new configuration knobs that meet the constraints each time they are changed.
(2) Figure \ref{fig:exp:2:subfig:power} shows the normalized power for each frame. 
Initially, when the QoS constraint is 10, the memory power is around \SI{80}{\%}. 

When the QoS constraint is relaxed to 85 at frame 30, \seams{} can lower the memory energy consumption by finding a new configuration of knobs and keeps operating in that region until the constraint changes again. Overall, the normalized energy/frame required by \seams{} is \SI{75.9}{\%}.
Therefore we demonstrated that \seams{} is capable of (1) self-adapting to new quality constraints specified by applications through coordination, and (2) continuously converging on optimal configurations.

\FloatBarrier
\subsection{Additional workloads}
Figures \ref{fig:exp:subfig:kmns:qos} and \ref{fig:exp:subfig:kmns:power} show \seams{}' result for \texttt{kmeans} with a normalized energy/frame of \SI{76.4}{\%}.
Figures \ref{fig:exp:subfig:blk:qos} and \ref{fig:exp:subfig:blk:power} show \seams{}' result for \texttt{blackscholes} with a normalized energy/frame of \SI{70.9}{\%}. We observe that even though the interdependent dynamics between all three layers of memory hierarchy and the achievable QoS/power are complex and application-dependent,  \seams{} is able to meet the dynamic quality constraints by continuously finding new configurations corresponding to the new system goals.

\subsection{\seams{} Overhead}
\label{sec:overhead}

\begin{figure}[tbh]
  \centering
  \resizebox{\linewidth}{!}{
    \input{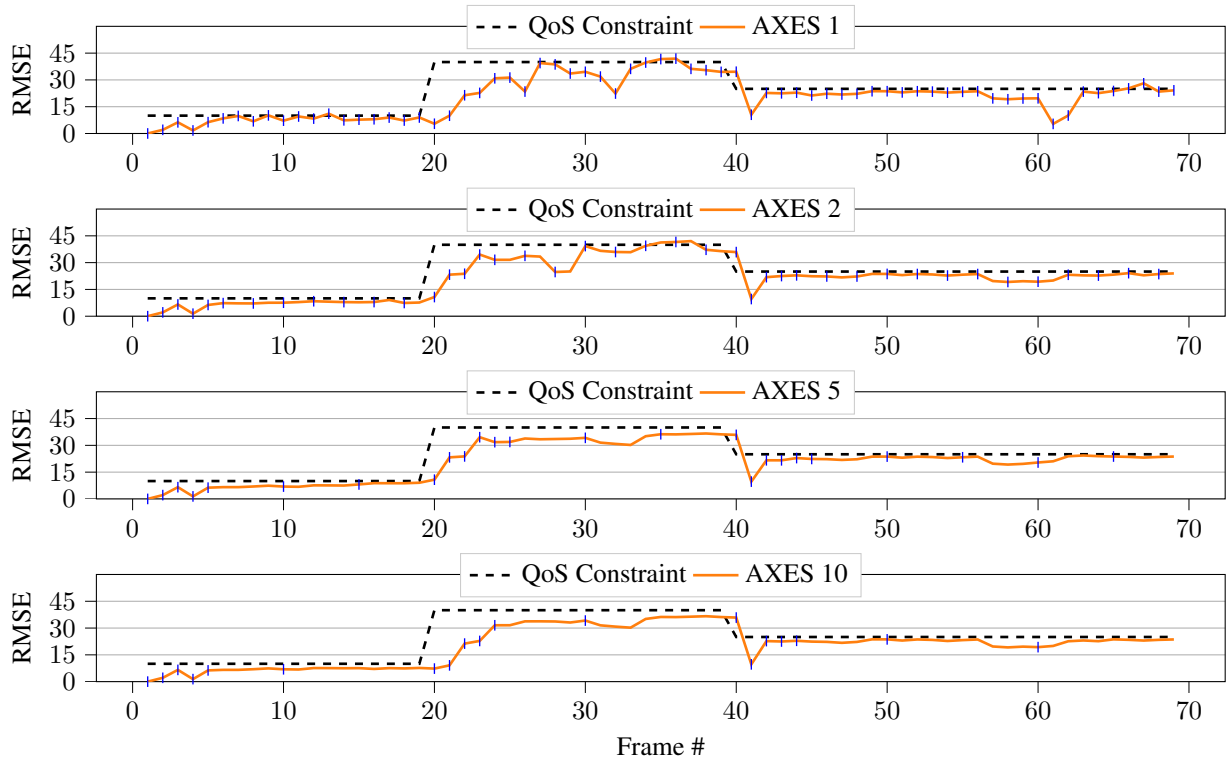}
  }
  \caption{\seams{} QoS at different invocation intervals for \texttt{k-means}. Blue ticks indicate  evaluation instances.}
  \label{fig:exp:overhead:intervals}
\end{figure}

All runtime approximation strategies suffer from two primary sources of overhead: 
(1) calculating the QoS value, and (2) runtime management. 
As mentioned in Section \ref{experimentalSetup_app}, to reap the benefits of approximation, \seams{} is not invoked for every input once the Q-values have been populated. 
In most cases, the quality monitor (e.g., the errors between pixels in canny edge detection) is embarrassingly parallel. 
If an additional core is available, the ground truth comparison can be performed in parallel. 
If \seams{} is invoked too frequently, the compute and energy overhead of the ground-truth comparison would not be justified. 
In Figure \ref{fig:exp:overhead:intervals}, we compare the QoS of k-means application at different intervals of invocation (marked with a blue tick). Whenever the system goals change, the subsequent five frames are always updated to adapt to the new setting. We observe that \seams{} can adjust to new goals with reduced invocation periods, at the risk of potentially missing self-optimization opportunities between invocations.
Even ignoring regular invocation completely, \seams{} can be used in an event-driven manner (e.g., updated at design time when a new system is developed, at runtime when a new application is available for approximation, or when the goals (QoS constraints) change). State-of-the-art alternatives do not self-optimize for these situations for a full memory hierarchy.

\begin{figure}[tbh]
  \centering
\begin{tikzpicture}

\definecolor{color0}{rgb}{0.12156862745098,0.466666666666667,0.705882352941177}
\definecolor{color1}{rgb}{1,0.498039215686275,0.0549019607843137}

\begin{axis}[
height=4cm,
legend cell align={left},
legend columns=2,
legend style={fill opacity=0.8, draw opacity=1, text opacity=1, draw=white!80!black, font=\small, at={(0.5,1.2)}, anchor=north},
minor xtick={},
minor ytick={},
tick align=outside,
tick pos=left,
width=0.7\linewidth,
x grid style={white!69.0196078431373!black},
xlabel={Invocation Period (\# frames)},
xmin=-0.5, xmax=3.5,
xtick style={color=black},
xtick={0,1,2,3},
xticklabels={1,2,5,10},
y grid style={white!69.0196078431373!black},
ylabel={\(\displaystyle \%\) of Baseline},
ymajorgrids,
ymin=0, ymax=30,
ytick style={color=black},
ytick={0,5,10,15,20,25,30}
]
\draw[draw=none,fill=color0,line width=1pt,postaction={pattern=bricks}] (axis cs:-0.25,0) rectangle (axis cs:0,21.349814008264);
\addlegendimage{ybar,ybar legend,draw=none,fill=color0,line width=1pt,postaction={pattern=bricks}};
\addlegendentry{compute overhead}

\draw[draw=none,fill=color0,line width=1pt,postaction={pattern=bricks}] (axis cs:0.75,0) rectangle (axis cs:1,10.674907004132);
\draw[draw=none,fill=color0,line width=1pt,postaction={pattern=bricks}] (axis cs:1.75,0) rectangle (axis cs:2,4.2699628016528);
\draw[draw=none,fill=color0,line width=1pt,postaction={pattern=bricks}] (axis cs:2.75,0) rectangle (axis cs:3,2.1349814008264);
\draw[draw=none,fill=color1,line width=1pt,postaction={pattern=dots}] (axis cs:0,0) rectangle (axis cs:0.25,0);
\addlegendimage{ybar,ybar legend,draw=none,fill=color1,line width=1pt,postaction={pattern=dots}};
\addlegendentry{power reduction}

\draw[draw=none,fill=color1,line width=1pt,postaction={pattern=dots}] (axis cs:1,0) rectangle (axis cs:1.25,10);
\draw[draw=none,fill=color1,line width=1pt,postaction={pattern=dots}] (axis cs:2,0) rectangle (axis cs:2.25,16);
\draw[draw=none,fill=color1,line width=1pt,postaction={pattern=dots}] (axis cs:3,0) rectangle (axis cs:3.25,18);
\end{axis}

\end{tikzpicture}
  \caption{\seams{} overhead for different intervals}
  \label{fig:exp:overhead}
\end{figure}
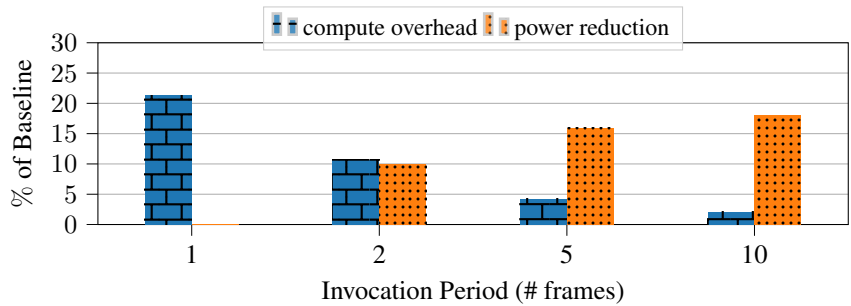

In Figure \ref{fig:exp:overhead}, we compare the compute overhead and the energy savings of \seams{} for various invocation periods based on the number of frames. 
Based on these observations, we invoke \seams{} every five frames in all of our evaluations.

The overheads of the current version of \seams{} are not preventative for the presented architecture and execution scenario.
However, the investigated architecture is unicore -- in a many-core system, the proposed control structure will not scale well due to configuration complexity if a single agent is responsible for configuring the entire memory system. 

Similarly, the current reward calculation is based on the quality reports of a single application utilizing the approximate memory segments -- multiple QoS applications running concurrently sharing approximate segments will complicate the agent. 
We believe a hierarchical or multi-agent architecture would effectively tackle the challenges of more complex systems and are topics for future investigation.

\FloatBarrier
\section{Conclusion}
In this paper, we address the challenge of designing a technology-agnostic runtime manager for approximate memory hierarchies. 
To this end, we propose \seams{}, the first self-optimizing model-free runtime manager for tuning approximation knobs in a unified multi-layer memory hierarchy to achieve acceptable QoS for diverse workloads while minimizing energy consumption. 
\seams{} uses temporal difference (TD) learning to learn directly from experience without a model of the environment's dynamics.
We develop an experimental case-study to evaluate \seams{} on a modified RISC-V processing core. We demonstrate the efficacy of \seams{} for configuration knobs when: (1) there is no quality constraint, and the system objective is to minimize power, (2) there is a quality constraint, and unknown inputs require reconfiguration of the knobs, and (3) the user dynamically changes the QoS constraints. 
\seams{} can automatically control approximate memory hierarchies regardless of memory technology or application. Due to these advantages, we believe that \seams{} has the potential for developing extremely energy-efficient systems across memory devices, parameters, and technologies.

\textbf{Acknowledgements} This work was partially supported by NSF grant CCF-1704859.
\bibliographystyle{unsrt}
\bibliography{main-arxiv}

\end{document}